\documentclass[aip,amsmath,amssymb,reprint]{revtex4-2}

\usepackage{caption}
\usepackage{subcaption}
\usepackage{float}

\usepackage{graphicx}
\usepackage{dcolumn}
\usepackage{bm}

\usepackage[utf8]{inputenc}
\usepackage[T1]{fontenc}
\usepackage{mathptmx}
\usepackage{etoolbox}
\usepackage{xcolor}
\usepackage{tabularx}
\newcolumntype{C}{>{\Centering\arraybackslash}X} 
\usepackage{hyperref}
\usepackage{draftwatermark}
\SetWatermarkText{PREPRINT}
\SetWatermarkScale{1}

\begin{document}


\title{Magnetic domain walls : Types, processes and applications}
\author{G. Venkat}
 \affiliation{Department of Materials Science and Engineering,  University of Sheffield,  S1 3JD,  UK}
 
\author{D. A. Allwood}
\affiliation{Department of Materials Science and Engineering,  University of Sheffield,  S1 3JD,  UK}

\author{T. J. Hayward}
\affiliation{Department of Materials Science and Engineering,  University of Sheffield,  S1 3JD,  UK}

\begin{abstract}
Domain walls (DWs) in magnetic nanowires are promising candidates for a variety of applications including Boolean/unconventional logic, memories, in-memory computing as well as magnetic sensors and biomagnetic implementations. They show rich physical behaviour and are controllable using a number of methods including magnetic fields, charge and spin currents and spin-orbit torques. In this review, we detail types of domain walls in ferromagnetic nanowires and describe processes of manipulating their state. We look at the state of the art of DW applications and give our take on the their current status, technological feasibility and challenges.  
\end{abstract}

\maketitle

\section{Introduction}\label{S1}

\noindent In recent years, there has been an explosive growth in research into magnetic nanostructures for memories, sensors and computation. Advantages of magnetic systems include their inherent non-volatility, fast speed of operation, low power consumption, nonlinear responses, negligible current leakage, and compatibility with CMOS fabrication techniques \cite{Vedmedenko2020roadmap}. This has led to magnetic random access memory (MRAM) products being available in the market as well as multiple demonstrations and proposals for Boolean and unconventional computing applications.

In this context, magnetic domain walls (DWs) in patterned magnetic nanowires have been proposed for many applications, which are introduced below. DWs are regions in a magnetic material between uniformly-magnetised domains through which magnetisation changes direction. DWs are typically small (with widths from a few nanometres to hundreds of nanometres), move relatively quickly (with reports of speeds of $18\,\rm{km/s}$) \cite{varga2007supersonic} and can be manipulated using a variety of stimuli, as we discuss later.

\noindent Magnetic DWs in nanowires have been developed as the information carriers for memory devices such as the switchable components of MRAM cells \cite{Khvalkovskiy2018magnetic} and shift registers, which include the racetrack memory architecture \cite{parkin2008magnetic} among others \cite{gaidis2011high, hayashi2008current}. These memories offer high storage densities with nanowire feature sizes scaled down to $20\,\rm{nm}$ \cite{fukami201320}, low read/write energies, low leakage current and years of data retention \cite{sun2018memory} in devices without movable parts. The intense research activities over years has now led to the prospect of commercialisation by exploiting CMOS fabrication and device/circuit level integration \cite{blasing2020magnetic, Everspin}.  

DWs in nanowires have also been proposed for logic applications \cite{allwood2005magnetic,luo2020current}. Control of the motion and interaction of DWs at nanowire junctions has allowed logic gates \cite{sun2018memory}, adders \cite{trinh2012domain,roohi2016tunable} and shift registers \cite{hayashi2008current} to be demonstrated. DWs also show complex behaviours when encountering energy barriers, such as at nanowire junctions. These behaviours are often stochastic in nature and highly sensitive to thermal perturbations, \cite{metaxas2007creep,hayward2015intrinsic} which was long regarded as a severe impediment to performing information processing tasks with magnetic DWs.  However, there have been examples of non-Boolean logic devices that make use of DW stochasticity to realise, for example, neuromorphic computing \cite{jin2019synaptic} and stochastic computing \cite{ma2018area} along with other devices such as a random number generator using DW-based Galton board \cite{sanz2021tunable}. The stochastic nature of DW depinning from nanowire junctions has also been shown to lead to reliable and complex many-body ensemble behaviour in large arrays of overlapping nanorings and proposed as the basis for reservoir computing  \cite{dawidek2021dynamically}.  


Beyond information technology applications, DWs in magnetic nanowires have been used to realise optical based \cite{wolfe1991fiber} and multiturn \cite{diegel2009new} magnetic field sensors. The stray magnetic fields from magnetic DWs in nanowires can also be used to interact with \cite{west2012realization} and position \cite{allwood2006mobile} ultracold atoms as well as control magnetic nanoparticles for biological sensing and cell positioning applications \cite{klein2013interaction}. There have been proposals \cite{bryan2010effect} and demonstrations \cite{bryan2010switchable,rapoport2017architecture} of trapping superparamagnetic beads using DW motion which can potentially be useful for manipulating biological cells and DNA. 

The movement of DWs in magnetic nanowires has been utilised for the majority of proposals of DW related technological applications and the last two decades have seen a surge in the study and manipulation of DWs using a variety of stimuli such as magnetic fields, charge and spin-polarised currents. Novel spin-related phenomena such as spin-orbit and spin-transfer torques have been utilised for controlling DW positions and velocities in nanowires in devices that move the technology from lab demonstrations to being closer to commercial realisation. 

This review discusses the various approaches of creating functionality from dynamics of DWs in both in-plane and out-of-plane magnetised nanowires, with a particular focus on the novel computing proposals that have emerged recently. We will also present fundamental aspects of the nature and control of DWs in nanowires to assist newcomers to the subject. 

The outline of the review is:
\begin{itemize}
    \item Section 2 details types of magnetic DWs patterned from thin films. This section also discusses available stimuli to manipulate and control magnetic DWs in patterned nanowires, including magnetic field, strain, voltage, and spin-polarised currents. We also discuss how various nanowire-based structures, including straight nanowires, nanorings, geometric defects, and nanowire junctions can be used to position DWs and tune their properties.
    \item Section 3 details DW processes in nanowires with particular attention to the stochastic features of these behaviours and the associated effects of thermal perturbations. 
    \item Section 4 concentrates on applications proposed using DWs in magnetic nanowires, including : memories, Boolean, neuromorphic and reservoir computing; and non-computing applications, including field sensors, biological applications, and transporting ultra-cold atoms.
    \item Section 5 concludes with a summary and future outlook.
\end{itemize}

\section{Magnetic domain walls}\label{S2}

\subsection{Domain wall structure}\label{S2-1}
Rectangular-cross-section magnetic nanowires typically have thickness from a few atoms to tens of nanometres, widths from 50 - 500 nm, and can extend in length to several tens or hundreds of micrometres. They are usually fabricated using standard lithography techniques such as electron beam lithography \cite{martin2002fabrication} or focussed ion beam milling \cite{basith2011direct} to pattern thin films into user-defined designs. We will be considering these rather than nanowires fabricated using electrodeposition techniques \cite{coey2001magnetic,schloerb2010magnetic} as more control on nanowire features is possible with lithography and the vast majority of applications proposed using DWs have considered lithographically patterned magnetic nanowires.

\begin{figure*}
   \centering
       \includegraphics[width=2\columnwidth]{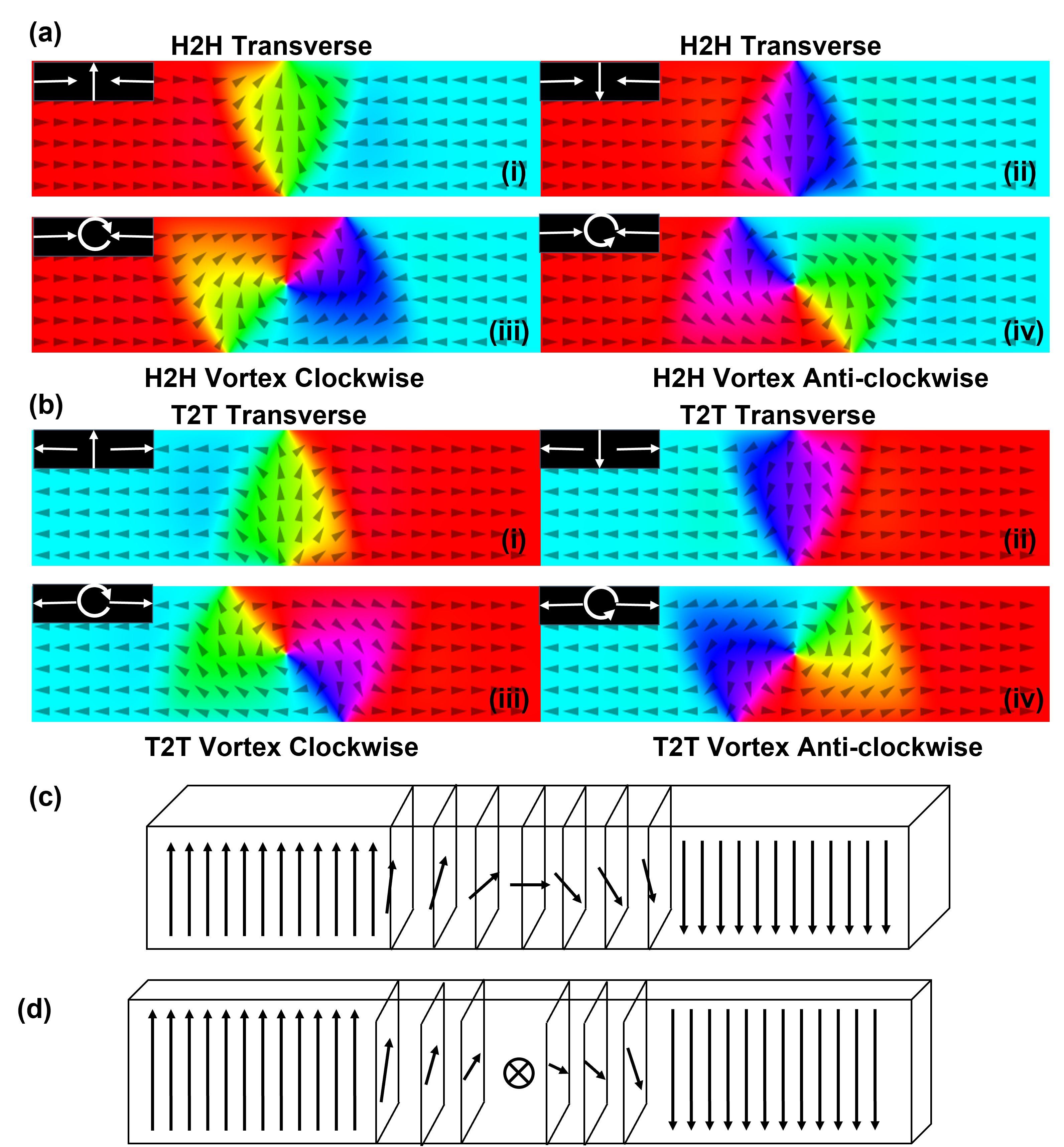}
\caption{(a) and (b): DW configurations in the top view of soft ferromagnetic patterned nanowires, obtained from micromagnetic simulations performed using Mumax3 \cite{vansteenkiste2014design}. Colour maps the local magnetisation direction and the inset to each figure part shows a simple schematic of the DW configuration. (c) and (d) shows schematics of N\'{e}el and Bloch type DWs in out-of-plane magnetised nanowires}\label{fig1}
\end{figure*}

Patterned nanowires made of soft magnetic materials such as $\rm{Ni}_{81}\rm{Fe}_{19}$ (Permalloy) have a magnetic structure dominated by shape anisotropy. The extended geometry of such a nanowire means magnetic domains are usually oriented in plane, along the wire length.  This results in DWs that lie directly across the width of a nanowire to separate oppositely-oriented magnetic domains. At their simplest, the possible configurations are referred to as  ‘head-to-head’ (H2H) when the DW separates domains with magnetisation oriented towards the wall and ‘tail-to-tail’ (T2T) when the domains are oriented away from the wall (Figure \ref{fig1} (a) and (b)). These DWs most often have in-plane magnetised (IM) configurations, due to the wires’ width usually being greater than their thickness. The simplest H2H (and T2T) DW structure is the ‘transverse’ configuration \cite{mcmichael1997head}, where moments in the centre of the wall are perpendicular to the nanowire edge (Figure \ref{fig1} (i) and (ii)). This occurs in thinner, narrower nanowires and results in decreased exchange energy and a high magnetostatic energy. In thicker, wider soft ferromagnetic nanowires, DWs form ‘vortex’ configurations where the moments rotate $360^{\circ}$ around a vortex core (Figure \ref{fig1} (iii) and (iv)). This structure has a low magnetostatic energy, at the expense of an increased exchange energy. DW structures with opposite chirality, i.e. the winding direction of magnetisation through the DW, are degenerate in a perfectly straight, defect-free nanowire, but lead to different behaviours when encountering asymmetric wire features or transverse magnetic fields \cite{bogart2008effect}. Asymmetric transverse DWs in soft ferromagnetic rectangular nanowires have also been predicted \cite{nakatani2005head} and then imaged using Fresnel-mode electron beam imaging \cite{petit2010magnetic}.

The in-plane DW textures shown in Figure \ref{fig1} can also be described using a system of magnetic topological defects \cite{tchernyshyov2005fractional}. For example, transverse DWs can be described in terms of magnetic edge ‘defect’ states with fractional winding numbers while vortex DWs can be described in terms of vortex 'defect' states with integer winding numbers. This approach provides a simple nomenclature for capturing DW structure and chirality and can be useful to categorise the behaviour of DWs and understand their interactions both with each other and with geometric features in nanowires. DW chirality has also been used functionally to manipulate DW trajectory \cite{pushp2013domain}, measure DW “fidelity” (a minimum length scale over which structural changes of the DW occur) \cite{lewis2009measuring}, and to form the basis of applications such as chirality-based DW memory cells \cite{atkinson2008controlling} and logic gates \cite{omari2014chirality}.

There has also been much interest in nanowires with perpendicular magnetic anisotropy (PMA) which are out-of-plane magnetised (OOPM). DWs in OOPM nanowires are usually much reduced in width of a few nanometres compared to in-plane-magnetised nanowires \cite{ravelosona2006domain}, which offers improved current-induced motion efficiency and higher data density \cite{tudu2017recent}. There are usually two types of DWs in such structures which are the N\'{e}el and Bloch types. These DWs usually have a Bloch structure in wider nanowires and transition to Néel DWs in narrower structures \cite{dejong2015analytic} (shown in Figure \ref{fig1} (c) and (d)). 




\subsection{Domain wall nucleation}\label{S2-2}

\begin{figure*}
   \centering
       \includegraphics[width=2\columnwidth]{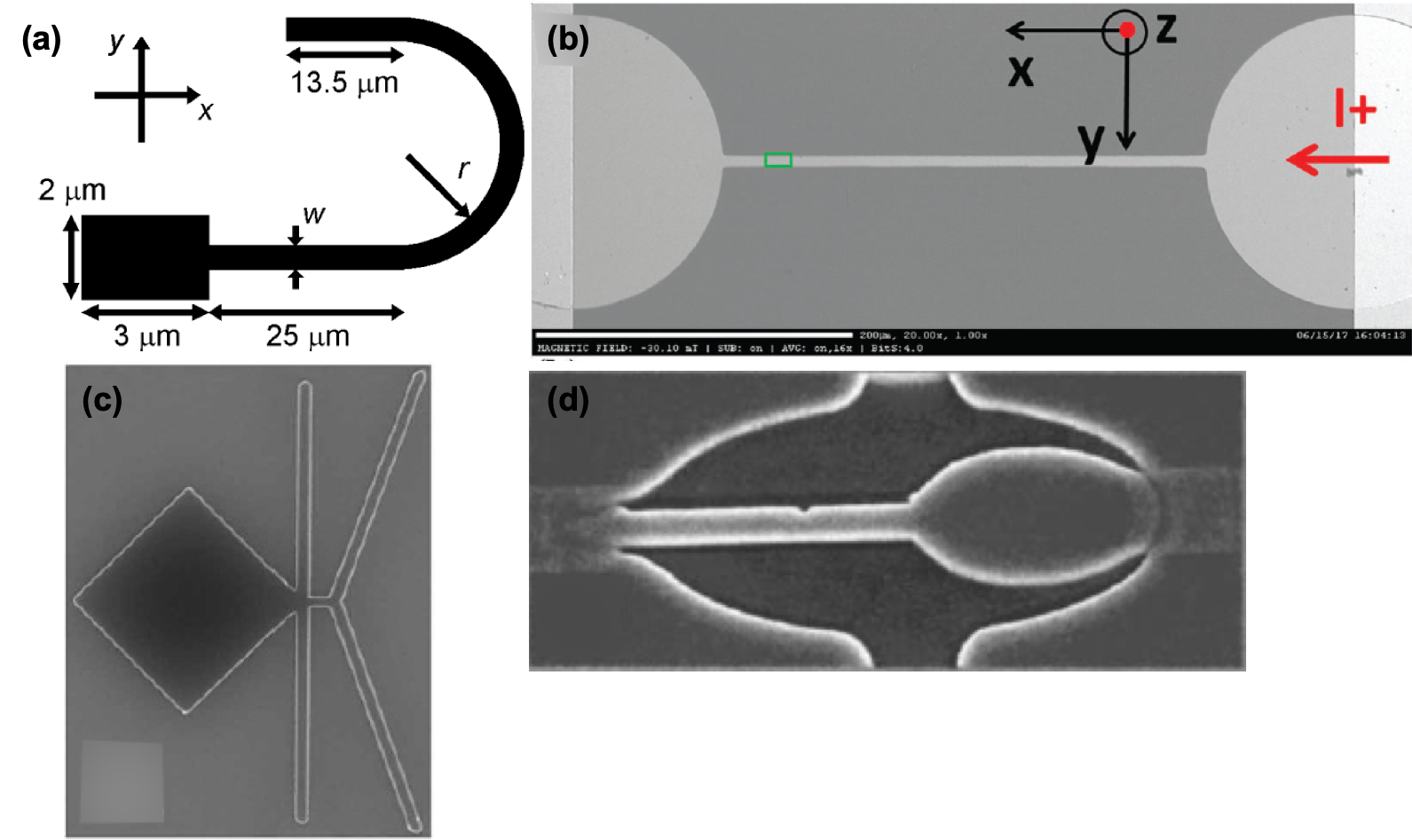}
\caption{Some of the different nucleation pads used for injecting DWs into nanowires are (a) rectangular \cite{bryan2009transverse}, (b) circular \cite{cui2018magnetization}, (c) attached \cite{goolaup2015transverse} and (d) elliptic \cite{thomas2005observation} in shape.}\label{fig2}
\end{figure*}

Controlling the nucleation of DWs in nanowires is important for many experiments and applications in order to test the behaviour of magnetic features or to represent data.  

The simplest approach involves fabricating a large nucleation pad, or ‘injection’ pad, at the end of a wire \cite{Atkinson2003magnetic}. Injection pads are typically made with lateral dimensions from single to tens of micrometres, and with a variety of shapes, e.g. rectangular, circular, attached or elliptical (shown in Figure \ref{fig2}). The weaker shape anisotropy of the pad compared with the nanowire means that an externally-applied global magnetic field causes a DW to nucleate within the pad first, and then be injected into the nanowire at magnetic fields far lower than the nucleation field of a simple wire end. 
Injection pads often lead to a range of DW structures being introduced to an adjoining IM nanowire as a variety of domain configurations may exist in the larger pad \cite{cowburn2002domain,bryan2006multimode}, although DW chirality can be controlled by careful design of an injection pad and placement of the attached nanowire \cite{mcgrouther2007controlled}.

DWs in IM nanowires can also be nucleated by using the Oersted field associated with currents driven through patterned current-carrying wires that cross over the magnetic nanowire \cite{koyama2011observation,fukami2011current}. This approach produces consistent DW structures using local excitations without the need for a global magnetic field, while the pulsed nature of the current allows more precise timing of nucleation events to synchronise with a wider experiment \cite{hayward2017beyond}.

\begin{figure*}
   \centering
       \includegraphics[width=2\columnwidth]{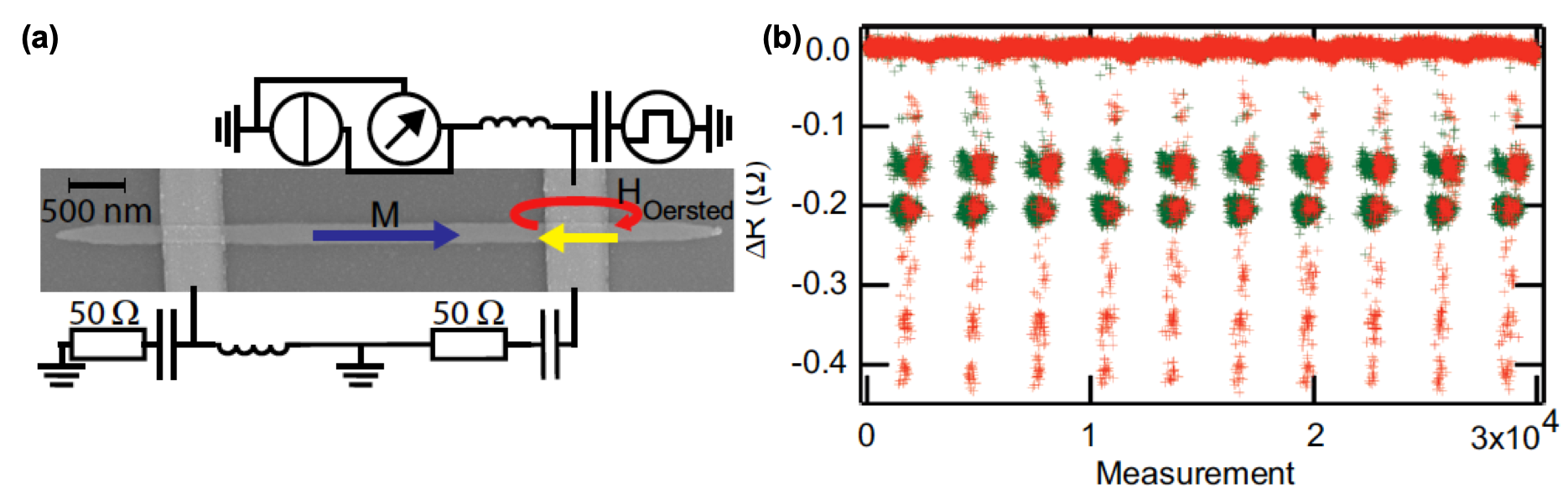}
\caption{A scanning electron microscopy image of the magnetic nanowire with a pair of vertically crossing current lines. (b) The variation of the nanowire resistance after one pulse (green) and after two pulses (red). Taken from [\onlinecite{stein2012generation}].}\label{fig3}
\end{figure*}

Stein et al. \cite{stein2012generation} have shown that combining current pulses of different polarities and external magnetic fields lead to reproducible creation and annihilation of DWs in an IM nanowire (shown in Figure \ref{fig3} (a)). Figure \ref{fig3} (b) shows changes in measured resistance after applying multiple current pulses and the different resistance levels corresponds to various types of DWs being formed in the nanowire. Specific choices of the polarity and duration of of the current pulses and polarity of magnetic fields led to stochastic pinning of DWs in the nanowire region bounded by the current lines as well as the injection of vortex and transverse walls into the nanowire. In the same year, Hayashi et al. \cite{hayashi2012microwave} used short lived current pulses (with $\sim\,\rm{ns}$ lifetimes) to decrease the magnetic fields required to nucleate DWs in IM nanowires. They found that the nucleation field reduced by around half and the distribution of the switching fields decreased in width as well and the authors attributed this to using localised fields generated by the current lines.  

Injections pads \cite{cayssol2004domain} and the Oersted fields from current carrying lines \cite{hayashi2006dependence} have also been used to inject DWs in OOPM nanowires. Zhang et al. \cite{zhang2016highly} used $\Pi$ shaped current lines for increased efficiency of DW injection in Co/Ni multilayer nanowires by modifying the shape of the current and localising the Oersted field to inject DWs. Other methods to inject DWs in OOPM nanowires include controlling the PMA by ion-irradiation \cite{franken2012shift} and ion milling \cite{tanigawa2008current} which change the physical structure and using Joule heating in particular regions of the nanowire via current injection \cite{narayanapillai2014thermally}.

\subsection{Domain wall motion}\label{S2-3}


There are several approaches to achieve controlled DW motion in patterned magnetic nanowires. It is well known that, in general, the application of a magnetic field causes ferromagnetic domains parallel to the field to grow and oppositely-oriented domains to shrink, with the changes in domain size mediated by DW motion. This can be understood at a more microscopic level by examining the Landau-Lifshitz-Gilbert (LLG) equation of motion \cite{gilbert2004phenomenological}
\begin{equation}
    \frac{\rm{d}\mathbf{M}}{\rm{d}t} = \mu_{0}\gamma\left(\mathbf{M}\times \mathbf{H}_{\rm{eff}} \right) + \frac{\alpha}{|\mathbf{M}|}\left(\mathbf{M}\times \frac{\rm{d}\mathbf{M}}{\rm{d}t} \right)\label{eq1}
\end{equation}
Here $\mathbf{H}_{\rm{eff}}$ is the effective local magnetic field that acts on the magnetisation, $\mathbf{M}$ in the time interval $\rm{d}t$. $\gamma$ is the electron gyromagnetic ratio, $\mu_{0}$ is the permeability of free space and $\alpha$ is the Gilbert damping parameter. $\mathbf{H}_{\rm{eff}}$ is calculated from the derivative of the system energy with changes in $\mathbf{M}$ \cite{hillebrands2003spin}, which includes any applied external field, $\mathbf{H}_{\rm{app}}$, as well as other energy contributions, e.g. magnetostatic, magnetocrystalline, and exchange energy terms.   Equation (\ref{eq1}) is most accurate when solved at a scale up to the exchange length of a material, i.e. the length over which magnetisation remains approximately uniform. 

The first term of the left-hand side of Equation (\ref{eq1}) shows that the initial response of $\mathbf{M}$ is to precess around $\mathbf{H}_{\rm{eff}}$. In IM nanowires and in a 1-D approximation, this means that a magnetic field applied parallel to the long axis of the nanowire (the longitudinal field component) will cause DW magnetisation, which is orthogonal to the field, to start to precess out of plane. This, in turn, creates an out-of-plane demagnetisation field that acts upon the DW’s remaining in-plane component of magnetisation and causes it to precess towards alignment with the applied field direction, expanding the magnetic domain in this process \cite{porter2004velocity}. The second term describes the damping of the precessional motion, which occurs in all ferromagnetic materials and, ultimately, causes local magnetisation to align to the local $\mathbf{H}_{\rm{eff}}$. Similar dynamics occur in OOPM materials with a magnetic field applied out of plane where the PMA plays a significant role. 



The fields required to propagate DWs are usually low ($1-20\,\rm{mT}$) both for soft magnetic materials, such as Permalloy \cite{beach2005dynamics} and for out-of-plane manipulation in some OOPM materials like Co/Pt \cite{emori2012time} although OOPM systems such as FePt/Pt require considerably higher fields \cite{attane2006thermally}. The magnetic fields are usually generated by an external electromagnet, and the resulting high power consumption usually limits the use of field-driven DW motion to laboratory experiments. There has, therefore, been intense interest in developing alternative, low-power methods of controlling DW motion and position.

DW motion in magnetic nanowires can also be induced by passing electric current directly through the wires. This approach is popular due to it offering local addressability, power efficiency, and fast DW motion, and has resulted in a large number of proposals for DW-based MRAMs and a commercial memory device \cite{fukami2009low}. Current control of magnetisation in thin films has been reviewed extensively from theoretical \cite{tatara2008microscopic} and experimental \cite{boulle2011current,beach2008current} viewpoints. Briefly, the spin angular momentum carried by a spin-polarised charge current causes an adiabatic ‘spin transfer’ torque (STT) \cite{berger1978low,zhang2004roles,thiaville2005micromagnetic} to be applied to the DW, which causes DW displacement. STT can also have a non-adiabatic contribution with an unclear origin with contenders being the force on the conduction electron due to gradient in the \emph{s-d} exchange field, linear momentum transfer and spin-flip scattering \cite{beach2008current}. STT is very relevant technologically as it can eliminate the need for magnetic fields and, therefore, for inefficient solenoids or electromagnets. STT has been proposed (in simulations) to nucleate DWs at specific locations in nanowires \cite{sbiaa2014multi} and Phung et al. \cite{phung2015highly} showed that they could inject DWs in nanowires with both IM and OOPM regions (obtained using ion implantation) using STT although the injection was stochastic in nature which might not be suitable for Boolean applications. One challenge with SST driven DW motion is that the position of the electrical contacts necessary for current-induced motion, of course, create natural limits to the extent of DW motion in nanowires \cite{klaui2005controlled}. Furthermore, the high current density for DW motion  ($10^{10}-10^{12}\,\rm{A/m^{2}}$) remains a technological challenge, both in terms of power efficiency and the resultant heating, which might damage the structure or change its behaviour. This is partly due to current-driven DW motion having an intrinsic pinning, even for an ideal wire, which leads to a threshold current both for IM \cite{beach2008current} and OOPM systems \cite{koyama2011observation}. The result is that sustained STT driven DW motion usually requires an external magnetic field \cite{yang2007spin,beach2006nonlinear,hayashi2006influence,emori2011roles}. Various proposals have been made for decreasing the current densities required for efficient control of DW motion, including using changing aspect ratios of nanowires \cite{yamaguchi2006reduction}, different geometries of the current injection \cite{boone2010rapid}, and OOPM materials \cite{fukami2008micromagnetic,jung2008current,komine2011reduction}. Tailoring the PMA in OOPM systems have been predicted lead to increased efficiency of SST driven DW motion \cite{emori2011enhanced}. There have also been interesting reports of lower switching current densities in systems involving synthetic ferrimagnets \cite{lepadatu2017synthetic} and antiferromagnets \cite{yang2015domain}, due to the in-plane magnetic interactions in them leading to narrow DWs.

\begin{figure*}
   \centering
       \includegraphics[width=2\columnwidth]{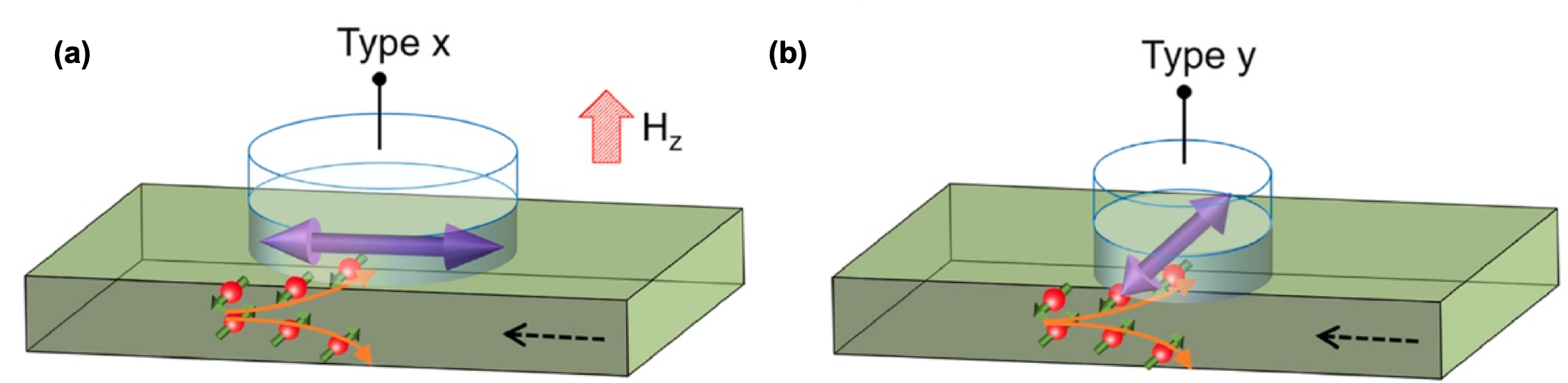}
\caption{Types (a) $x$ and (b) $y$ configurations for the SOT due where the injected current (black dashed arrow) and magnetisation (purple arrow) lie along and perpendicular to each other respectively. The splitting of electrons of different spin are shown as well. The 'type-x' configuration required an out-of-plane for breaking the symmetry and switching the magnetisaton. Taken from [\onlinecite{song2021spin}].}\label{fig4}
\end{figure*}

Recently, non-magnetic (NM)/ferromagnetic (FM) bi-layers have opened up multiple schemes of controlling the magnetisation in the magnetic layer. The spin Hall effect (SHE) \cite{dyakonov1971current} describes the generation of a transverse spin current at the FM/NM interface due to spin-orbit coupling in the NM. The SHE can significantly affect the DW dynamics in nanowires \cite{seo2012current} and can even reduce the current densities required for DW switching \cite{ryu2013current,haazen2013domain}. The torque exerted by this transverse spin current on the magnetisation in the FM is known as spin-orbit torque (SOT) and this has been studied intensely  in the last few years \cite{ramaswamy2018recent,song2021spin}. Depending on the relative orientation of the injected current and the magnetisation in the FM, there can be two configurations of SOT for switching as shown in Figure \ref{fig4}. The "type-x" configuration require a magnetic field out-of-plane of the nanowire for breaking the symmetry of the system for deterministic switching which can be inconvenient for technological applications. SOT-induced schemes overcome limitations of STT with the added advantages of reduced power consumption and after device operation. SOT-induced DW depinning has been predicted \cite{zhang2017spin} and demonstrated \cite{lo2014spin} in Ta/CoFeB thin films nanowires, making available an added method of controlling DWs \cite{yoon2017anomalous}. While most such studies have been with OOPM systems, there have also been analytic predictions of SOT driven DW motion in IM systems such as Pt/CoFeB/MgO \cite{kohno2020domain}. Further novel phenomena such as the Rashba effect and the Dzyaloshinskii-Moriya interaction (DMI) can also be used to tune DW velocity \cite{miron2011fast,martinez2013current,ajejas2017tuning} by controlling the DW energy using these effects. For instance, Dao et al. \cite{dao2019chiral} have shown that DWs can be controllably pushed into nanowires using the chiral interactions induced by DMI between in-plane and out-of-plane magnetised trilayers of $\rm{Pt}/\rm{Co}/\rm{AlO}_{x}$ in combination with SOTs induced using currents. The capability of SOT to nucleate DWs are found to strongly be chirality dependent and can required current densities less than traditional current induced DW injection methods. SOT offers a novel and useful control of DWs using electric currents for device applications, although some challenges such as efficient magnetisation switching without out-of-plane fields have to be overcome.

\begin{figure*}
   \centering
       \includegraphics[width=2\columnwidth]{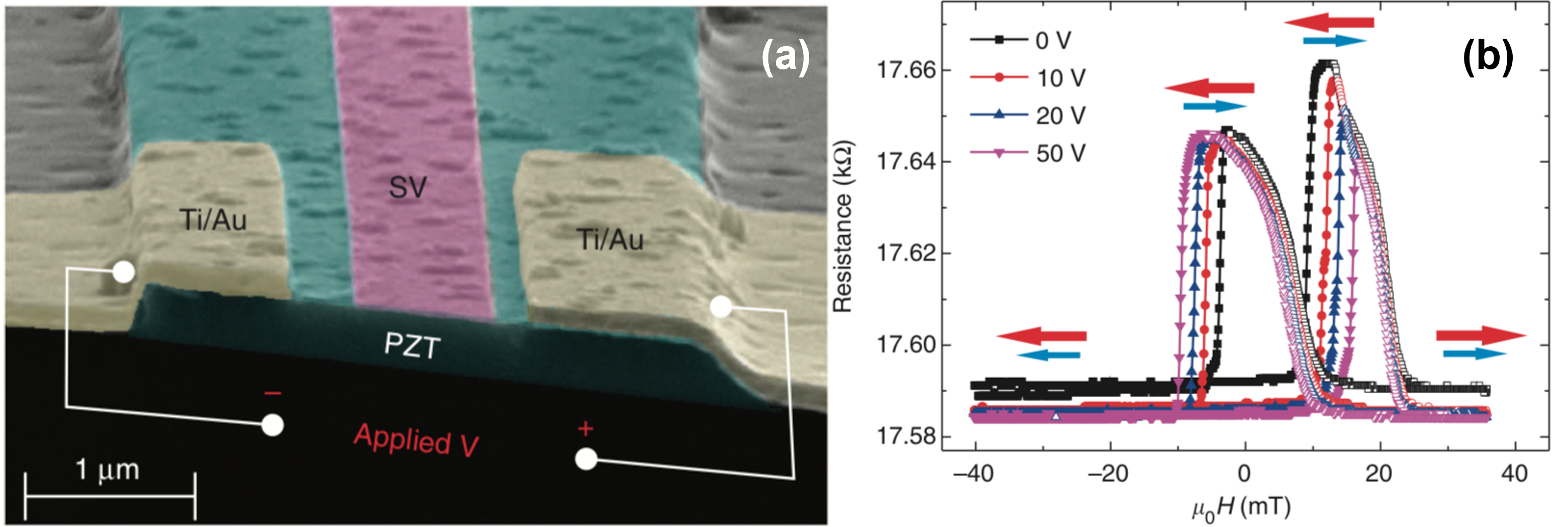}
\caption{A scanning electron microscopy image of the piezoelectric/multilayer device used for showing effects of strain on the magnetic response.  (b) The variation of the giant magnetoresistance of the device for different voltages applied to the piezoelectric which leads to a modification of the strain on the nanowire. Taken from [\onlinecite{lei2013strain}].}\label{fig5}
\end{figure*}

\begin{figure*}
   \centering
       \includegraphics[width=2\columnwidth]{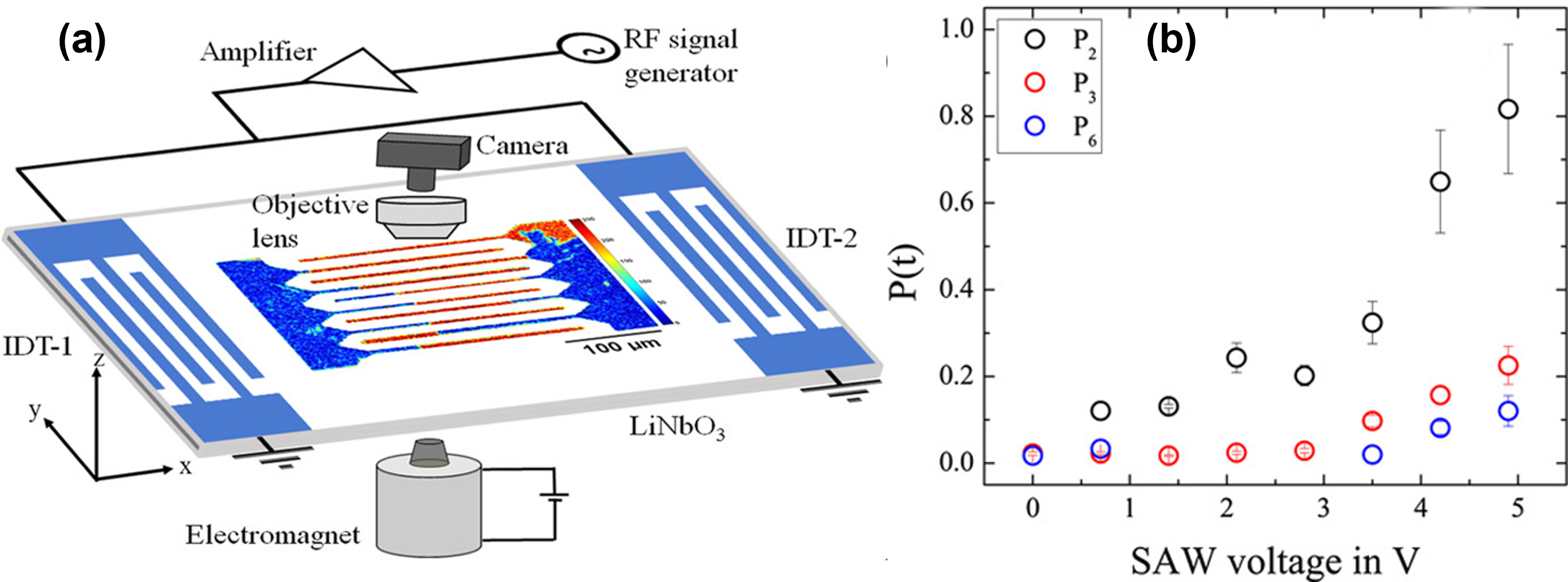} 
\caption{A schematic of the setup used to modify DW pinning/depinning using SAWs showing the magnetic stripes and integrated digital transducers (IDTs). (b) The depinning probability as a function of increasing SAW voltage for three representative pinning sites. Taken from [\onlinecite{adhikari2021surface}].}\label{fig6}
\end{figure*}

DWs in nanowires can also be controlled using applied strain in magnetostrictive materials, in which strain mediates magnetisation changes via inverse magnetostriction, also called the ‘Villari effect’ \cite{hristoforou2007magnetostriction}. Metallic magnetostrictive nanostructures are often coupled to underlying piezoelectric materials to allow strain in the piezoelectric to be created via electric potentials applied to patterned contacts before being transmitted to the magnetic element (Figure \ref{fig5} (a)).  Such systems are termed ‘artificial multiferroics’, due to their heterogeneous structure and show a variety of magnetic and DW driven phenomena \cite{meier2015functional}. Local variations in static strain across the length of a magnetostrictive nanowire create potential barriers or wells that can be used to addressably pin or even move DWs along the wire length \cite{dean2011stress,yu2022strain}. Lei et al. \cite{lei2013strain} have shown that by applying a static strain in piezoelectric/multilayer magnetic structures, the magnetic response can be reproducibly tuned (Figure \ref{fig5} (b)). Other proposals for manipulating DWs using strain include applying dynamic strain profiles and Dean et al. \cite{dean2015sound} proposed using standing surface acoustic waves (SAWs) along the length of an IM magnetic nanowire to create an array of DW pinning sites. Subsequently, Adhikari et al. \cite{adhikari2021surface} found that exciting DWs in an array of OOPM Co/Pt multilayer nanowires using SAWs (created using integrated digital transducers and shown in Figure \ref{fig6} (a)) resulted in an increase in the DW depinning probabilities by approximately a factor of 10 (Figure \ref{fig6} (b)). Technological proposals based upon strain-induced control of DWs include memories \cite{novosad2000novel} and manipulation of magnetic beads in a fluid \cite{sohn2015electrically}.

Optical pump-probe techniques have also been used to move DWs in IM Co/Ni films \cite{sandig2016movement} and open up possibilities of ultrafast control and writing of magnetisation. We therefore see that many varied and technologically relevant schemes are available for initiating and controlling DW states in both IM and OOPM nanowires. 

\subsection{Positioning domain walls}\label{S2-4}

Geometric modifications have been also used to create sites for pinning/depinning of DWs in nanowires by altering the magnetic energy landscape. In magnetic nanowires, local geometric modifications invariably cause change to the magnetostatic energy of a DW, with subsequent rearrangements in the magnetic configuration of a DW at the modified side leading to changes in exchange energy \cite{bogart2009dependence}. These changes depend upon the structure and chirality of the DW, and the precise geometry of the modified region, which has given a lot of scope for investigation.

\begin{figure*}
   \centering
       \includegraphics[width=2\columnwidth]{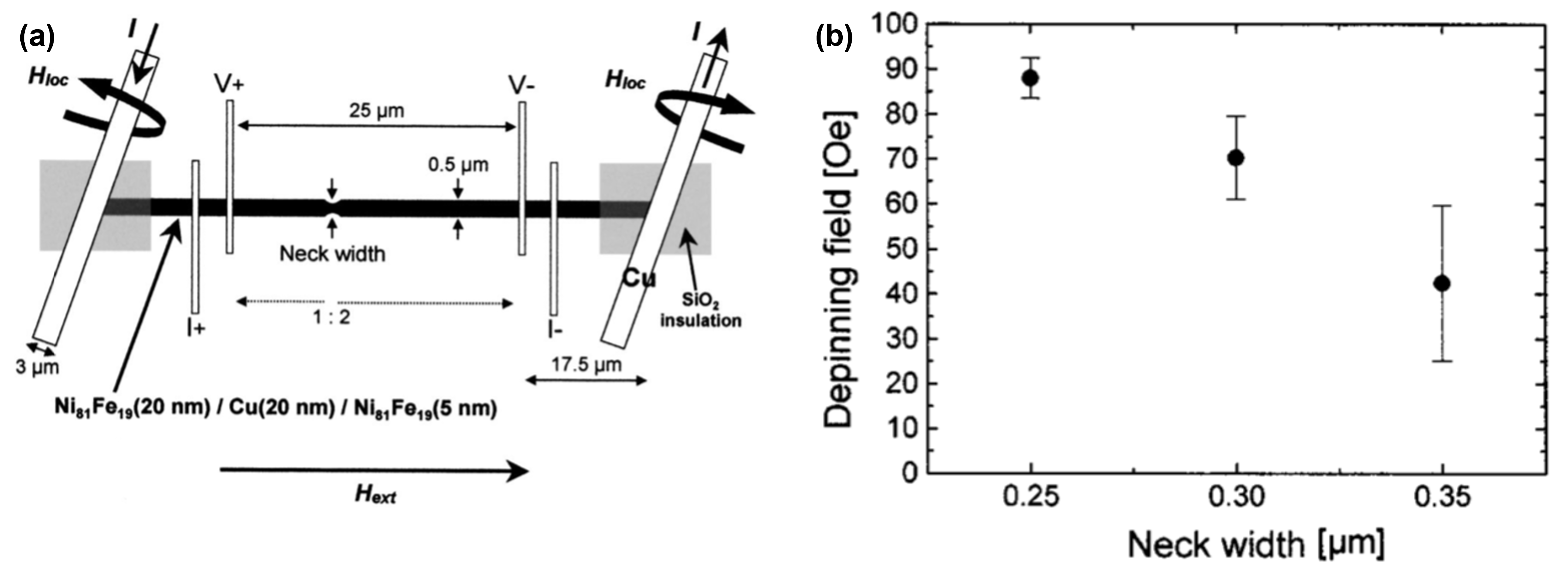}
\caption{A schematic of a trilayer nanowire (in black) and DWs in them were manipulated using Oersted fields created by passing currents through Cu wires shown at the ends of the nanowire. (b) The variation of the depinning field of a DW from the notch with the neck width of the notch. A higher depinning field is observed for a notch with lower neck width. Taken from [\onlinecite{himeno2003dynamics}].}\label{fig7}
\end{figure*}

The nucleation and pinning of DWs in nanowires are dependent on the defects introduced during fabrication of the wire. Dutta et al. \cite{dutta2017spatial} found that edge roughness of the wire can trap DWs at length scales less than the resolution of the fabrication process both in IMA and PMA nanowires. They studied a system of nanowires in the regime where the DW width was smaller than the correlation length of the wire roughness (i.e. the length scale defining the transition between smooth and rough length scales of the nanowire) and found a discrete distribution of DW traps in the nanowires which limits the precision of the placement of a DW in the nanowire. 

Himeno et al. \cite{himeno2003dynamics} used artificial necks of varying width in IM NiFe/Cu/NiFe trilayer submicrometer wires to control the depinning fields of the DWs (Figure \ref{fig7}). Kl{\"a}ui et al. \cite{klaui2003domain} also used notches in an IM curved ring nanowire and by injecting currents between various contacts positioned between the notches, they found multiple transitions in the anisotropic magnetoresistance (AMR) response from the ring which corresponded to DW pinning at the notches. Such studies led to further investigations of notches of different shapes \cite{bogart2008effect,bogart2009dependence}, modifying the position of notches in the nanowire \cite{huang2009domain}, and proposals of stepped nanowires for memory applications \cite{al2019staggered}. These demonstrated in various ways how the potential landscape and depinning fields of DWs could be modified. 

Asymmetric geometric defects can be used to create additional levels of control in IM nanowires. DWs of opposite chirality generally interact differently with defects that are present on one side of a nanowire only, regardless of whether they are notches \cite{petit2008domain,eastwood2011chirality}, protuberances \cite{bryan2007symmetric}, or adjoining wires at $90^{\circ}$ to the original wire \cite{lewis2009measuring}. This has been used as a means of selecting one chirality over another \cite{goolaup2011dependence} and the basis of a logic system using vortex DWs \cite{omari2014chirality,omari2019toward}. In all cases, if the geometric modification is repeated on the other side of the wire, the chirality-filtering is removed and DWs of the same structural type pass through with identical depinning fields \cite{hayward2017beyond}. Asymmetry along the wire direction of a notch or protuberance shape creates a different energy pathway for DWs travelling through the wire in opposite directions. This leads to depinning fields that depend on the direction a DW is travelling, in effect creating a ‘DW diode’ using both in-plane \cite{allwood2004domain, himeno2005propagation} and OOPM materials \cite{luo2021field}.

External stimuli can be used to control DW pinning at geometric features. For example, a DW in an IM nanowire can be pinned through magnetostatic interactions with an adjacent and close-lying magnetic nanowire \cite{hayward2010pinning,hayward2010direct,o2009near,o2011dynamic,purnama2011current}. Also, an externally-applied transverse magnetic field can be used to tune the DW depinning probability function at notches \cite{omari2014chirality}, which gives a greater degree of control. Kim et al. \cite{kim2009depinning} studied the depinning of DWs at notches of different gap sizes in an OOPM nanowire and demonstrated that the depinning fields and times could be controlled by changing the gap sizes. The Oersted field from current-carrying lines patterned across magnetic nanowires can also be used to pin DWs in an addressable manner \cite{jang2009current,jang2012current,hayashi2006dependence}, although the Oersted fields created using these lines are usually relatively low. 

Precise understanding of energy pathways allows prediction of DWs experiencing potential wells or potential barriers. This can be tested experimentally by pinning a DW and then reversing the applied magnetic field. A potential barrier will lead to ‘pinned’ DWs moving back away from the defect site under low fields but a pinned DW will require a similarly-large field to move from the site in either direction \cite{hayward2010pinning}. However, DW depinning from geometrical modifications is complicated by thermally-activated stochastic processes \cite{hayward2015intrinsic}, which we will explore more later.

DWs have also been positioned by modifying the global wire geometry. Magnetic fields can cause DW motion in nanowires whenever there is a field component parallel to the wire direction. Curved nanowires, therefore, provide a natural limit to the extent of DW motion and they have been used to assist in exploring the nature of various DW phenomena in nanowires using magnetic fields and currents in both IP and OOPM systems\cite{ho2018field,vernier2004domain,garg2017dramatic}. An in-plane rotating magnetic field causes DWs to propagate around curved wire regions in the directions of same handedness as the field rotation, which has been exploited for clocking in logic \cite{allwood2005magnetic} and sensor \cite{diegel2009new,diegel2004360} systems. The depinning fields in curved nanowires is sensitive to the degree of curvature of the nanowire and Glathe et al. \cite{glathe2012magnetic} suggested that DW pinning in a curved IM nanowire can be better controlled by designing corners as a polygonal line. Modifying wire geometry can also be used to control DW chirality, with vortex DWs in curved IM nanowires subject to an externally-applied magnetic field observed to have chirality that is highly dependent upon the wire width and magnetic field direction \cite{schonke2020quantification}. On the introduction of width gradients in half rings to form structures having either point of axis symmetry, the authors could generate vortex DWs with either same or opposite chirality via different stochastic switching pathways. The stochasticity in these pathways could be controlled by modifying the width gradient and temperature and the authors indicated that this could be useful for applications.

\begin{figure*}
   \centering
       \includegraphics[width=2\columnwidth]{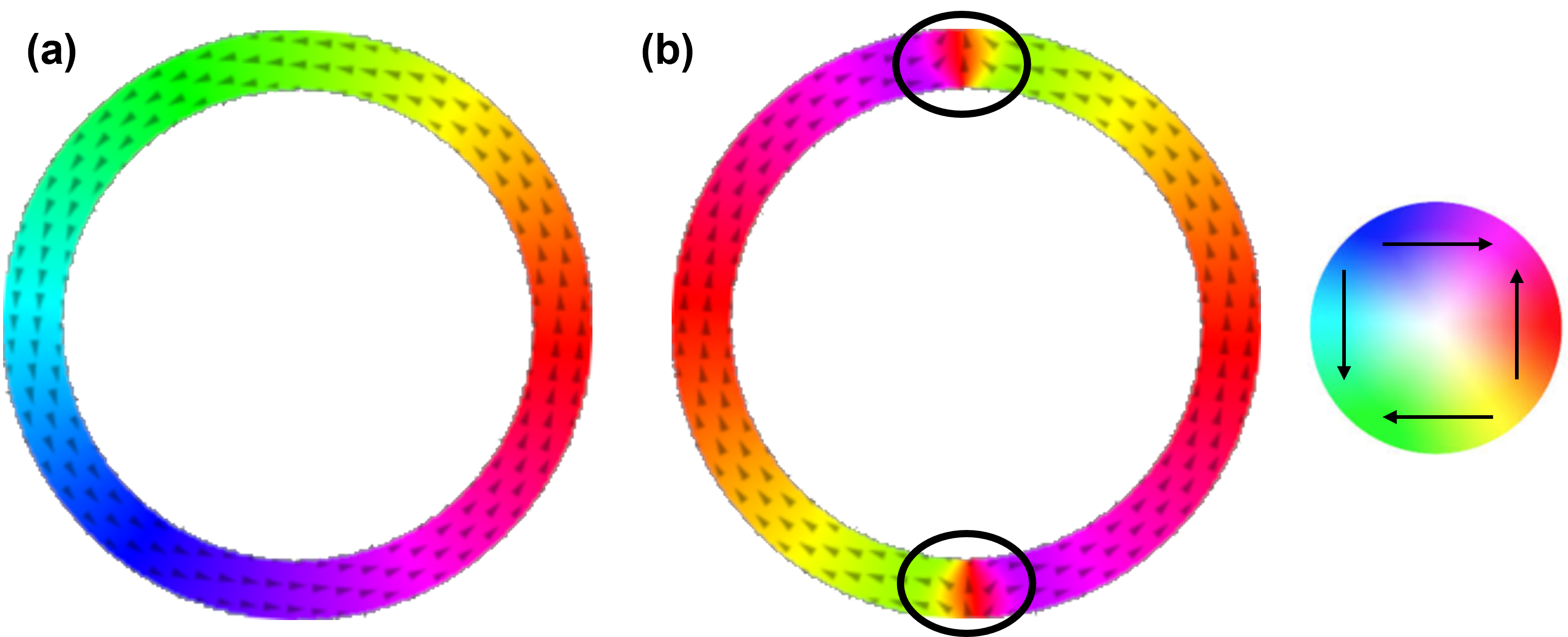}
\caption{(a) ‘Onion’ and (b) ‘vortex’ magnetic states in a soft ferromagnetic nanowire ring. The DWs in the ‘onion’ state are highlighted by the circles and the colour wheel maps the colours to magnetisation orientation.}\label{fig8}
\end{figure*}

IM ferromagnetic rings offer an interesting combination of having magnetic properties similar to a straight nanowire locally but with a wider curved geometry that creates well-defined magnetic states and makes positioning DW straightforward.  The lowest energy magnetic configuration of a soft ferromagnetic ring is the ‘vortex’ state, in which magnetisation is oriented circumferentially, i.e. following the local wire direction around the ring, and no DWs are present (Figure \ref{fig8} (a)). DWs can be introduced in pairs, with the so-called ‘onion’ state referring to a configuration with two DWs, one H2H and one T2T, on opposite sides of the ring (Figure \ref{fig8} (b)). Other states are possible \cite{castano2003metastable} but the vortex and onion configurations form the bulk of interest \cite{klaui2003vortex, vaz2007ferromagnetic}. Application of an in-plane magnetic field causes DWs to align on opposite sides of a ring in a radial direction parallel to the field direction \cite{klaui2003vortex}. This means in-plane rotating magnetic fields of sufficient strength can be used to cause the DW pairs of an onion state to propagate around the ring in the same direction as the field rotation \cite{negoita2013domain}. Bisig et al. \cite{bisig2015dynamic} reported that rotating fields can also be used to set the chirality of DWs with high fidelity in rings of varying diameters and width with thermal activation playing a role in the switching process.

The magnetic behaviour of ring arrays can be quite different from that of individual rings. For instance, the magnetostatic interactions within an array of close-lying rings can cause a significant difference in the switching fields of arrays with rings of varying widths and spacings \cite{klaui2005domain}. In fact, the switching of Cobalt rings arranged in a chain transverse to the switching field always occurs in pairs \cite{welp2003magnetization}. These additional complexities in the behaviour of multiple elements can be important for using such systems for non-Boolean computing architectures, as shall be described later on.

\begin{figure*}
   \centering
       \includegraphics[width=2\columnwidth]{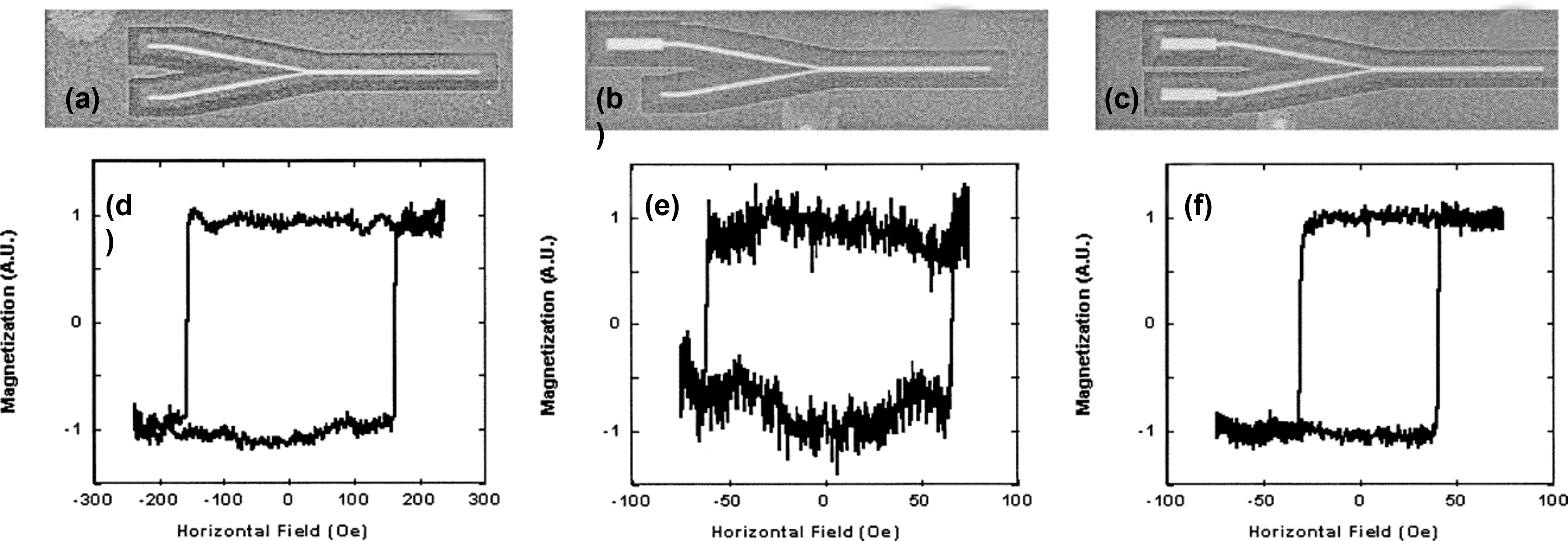}
\caption{FIB images of typical three-terminal structures fabricated, with (a) zero, (b) one, and (c) two DWs being injected into the junction. Directly below each image is a hysteresis loop of the output arm corresponding to the type of structure shown showing the difference in fields at which the magnetisation switches. Taken from [\onlinecite{faulkner2003controlled}]. }\label{fig9}
\end{figure*}

We shall now discuss the behaviour of DWs at junctions of nanowires. Faulkner et al. \cite{faulkner2003controlled} considered an IM three terminal nanowire junction of Permalloy for controlled DW injection (shown in Figure \ref{fig9} (a)-(c)). They found that the fields required for switching the magnetisation of the output arm of the device was significantly lower when DWs were injected from either one or both input arms compared to when no DW was injected in either input arm (seen from the switching fields in the hysteresis loops in Figure \ref{fig9} (d)-(f)). This was a consequence of a single DW (in the case of DW injection from one of the input arms) or linked DWs (when DWs are injected from both input arms) expanding in the output arm from the junction and causing the magnetisaton to switch. Pushp et al. \cite{pushp2013domain} subsequently considered an IM Y-shaped nanowire junction and could reliably inject DWs of a given chirality into each arm of the Y-shaped structure by varying the strength of the injection field. Such IM junction based nanowire devices were subsequently proposed for DW based Boolean logic operations and we shall describe these in Section \ref{S4-2}. DW behaviours at junctions have also been considered in OOPM systems and Kwon et al. \cite{kwon2017asymmetrical} considered propagation of N\'{e}el DWs in a bifurcated nanowire of multilayer Ni/Co grown on Pt. The DW injection into each of the two arms was at different applied fields and this was attributed to the DMI preferring one tilt direction of the DW surface leading to easier injection in one of the arms. We thus see that the behaviour of DWs at nanowire junctions offer additional opportunities of manipulating DW states for developing devices.

\section{Domain wall processes in nanowires}\label{S3}

DWs propagating in straight, patterned nanowires show a large array of dynamic behaviours over and above simple motion \cite{beach2005dynamics, mougin2007domain, tretiakov2008dynamics}. These behaviours are beyond stimulus driven deterministic DW evolution that was considered in the previous section and are heavily affected by thermally driven phenomena in the nanowires. We introduce these additional behaviours in brief below.

\begin{figure*}
   \centering
       \includegraphics[width=2\columnwidth]{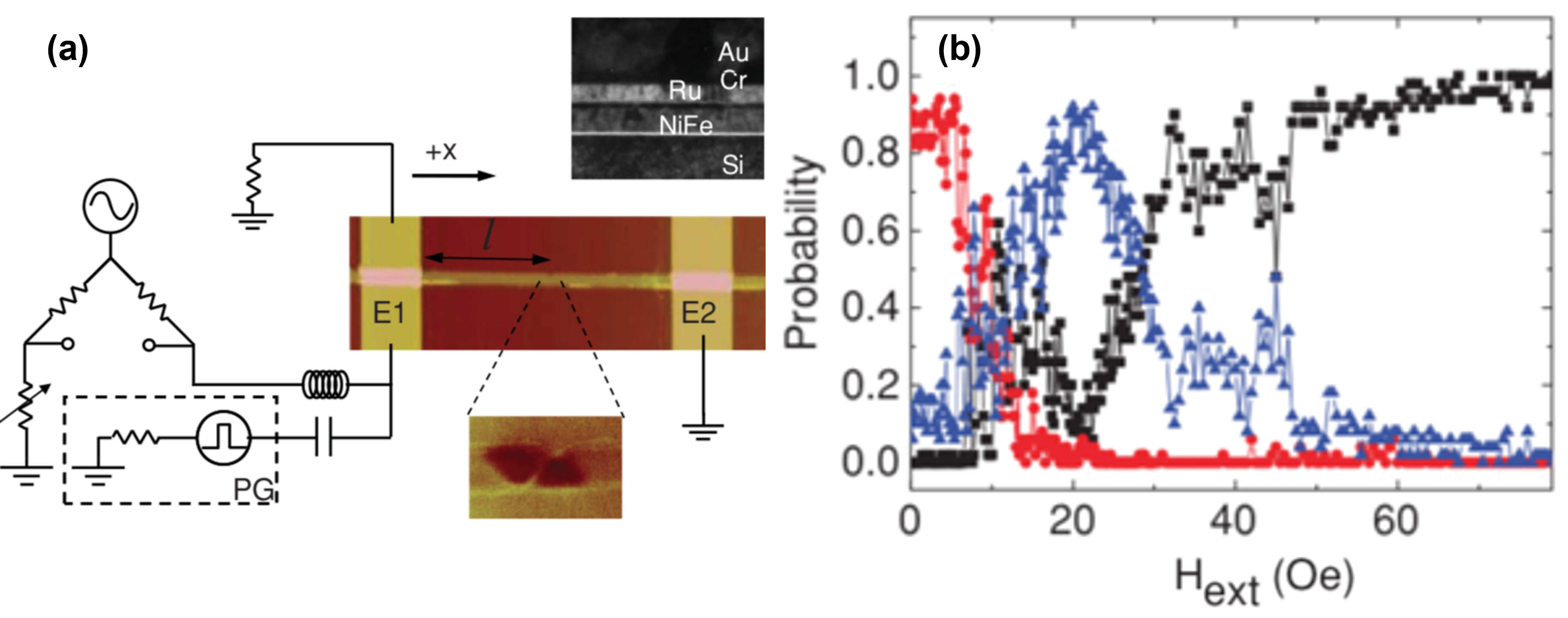}
\caption{(a) The measurement setup for a detecting the depinning of a DW in an IM notched nanowire. (b) The relative probability of a DW depinning from the notch (black), a vortex DW (red) and transverse DW (blue) being pinned at the notch. Taken from [\onlinecite{pi2011static}].}\label{fig10}
\end{figure*}

We refer to the paper by Hayward et al. for most of the discussions in the next two paragraphs \cite{hayward2015intrinsic}. DW processes are stochastic and experimental observations can be classified into three categories \cite{hayward2015intrinsic}:
\begin{itemize}
    \item DW propagation, even in defect free nanowires, is not deterministic even for magnetic fields greater than what is required for DW motion
    \item DWs pinned at engineered pinning sites have rich, multimodal depinning magnetic field distributions. In fact, it is common to observe the DW depinning probability increase sigmoidally with applied magnetic field. This was observed, for example, by Pi et al. \cite{pi2011static}, who studied DW pinning at a notch in an IM Permalloy nanowire (Figure \ref{fig10} (a)) by AMR (Figure \ref{fig10} (b) black line). 
    \item DWs move non-deterministically via defect sites when magnetic fields that would not be expected to be able to induce depinning are applied.
\end{itemize}

\begin{figure*}
   \centering
       \includegraphics[width=2\columnwidth]{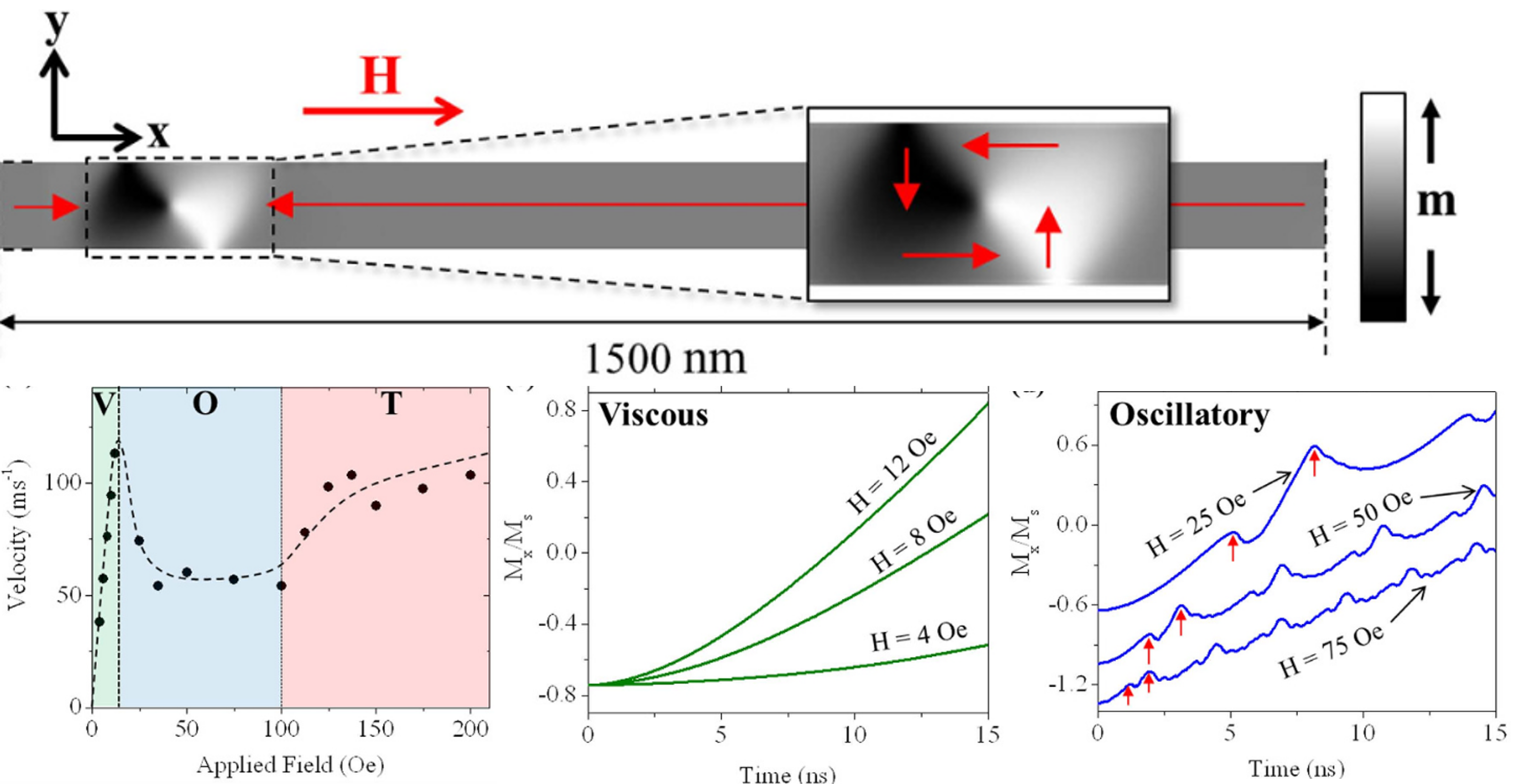}
\caption{(a) (Top) : A diagram showing the geometry of the nanowire simulated in [\onlinecite{hayward2015intrinsic}]. The initial vortex DW configuration is shown in higher detail. The plots below show the DW velocity and normalised magnetisation in different regimes of DW dynamics. Taken from [\onlinecite{hayward2015intrinsic}].}\label{fig11}
\end{figure*}

The simple picture of DW motion in a nanowire described in Section \ref{S2-3} holds above a threshold field strength, due to overcoming imperfections in a wire, and with constant mobility $\mu_{\rm{DW}} = \frac{v}{H}$, where $v$ is DW velocity and $H$ the driving magnetic field \cite{schryer1974motion,lee2007dynamic}. This ‘viscous’ regime remains until the ‘Walker breakdown’ field, the value of which depends upon nanowire material, geometry, and DW structure. Above this field, DWs first enter a regime of oscillatory motion and negative differential mobility, followed by a turbulent regime that sees the DW velocity mobility become positive once more \cite{hayward2015intrinsic} (refer to Figure \ref{fig11}). Walker breakdown processes see dramatic changes to the dynamic structure of DWs, accompanied by the near-periodic stalling of DW motion and transition of DW configuration between different types and chiralities. In the nanowire configuration considered in \cite{hayward2015intrinsic}, an oscillatory transition between VDW and TDW types of alternating chiralities occurs. This stochasticity of DW structure has been explained as thermally-driven fluctuations in magnetic field driven Walker breakdown pathways \cite{hayward2015intrinsic}. Furthermore, thermal perturbations affect these different regimes of DW dynamics quite significantly, especially above Walker breakdown, when the intrinsic instabilities of the DWs are heavily influenced by relatively small perturbations \cite{hayward2015intrinsic}. 

Such DW processes are not restricted to magnetic field stimuli. Similar dynamics to the field driven dynamics described above have been predicted and observed with current-driven DW motion in patterned nanowires. Duine et al. \cite{duine2007thermally} used the stochastic version of the LLG equation which has a stochastic field component to account for temperature driven perturbations. The authors found that the variation of the DW velocity with applied current was linear even without any STT at non-zero temperatures due to these perturbations indicating that they can play an important role. Subsequently, Hayashi et al. \cite{hayashi2006influence} found that current-driven DWs propagating through an IM Permalloy nanowire and pinned at a notch would have different transverse and vortex structure of different chiralities, despite identical experimental conditions and a single structure being used. They also found that each of the four DW states observed had substantially different depinning fields but similar depinning current densities and concluded that with current-induced depinning, all the DW types may be converting to the same DW state. Higher current densities again result in DWs experiencing Walker breakdown and turbulent dynamics, although the Walker threshold current density can be found above the point where the DW velocity has ceased to respond linearly to current \cite{mougin2007domain}. Hybrid approaches to driving DW motion have also been studied and have shown that application of a magnetic field lowers current driving thresholds \cite{koyama2012current,boulle2011current}. Furthermore, as seen in Section \ref{S2-4}, the patterned geometry of a magnetic nanowire plays an important role in determining DW dynamics. For example, modifying the curvature of nanowires has been predicted to modify the magnetic-field-driven DW velocities, precession, and oscillation frequencies \cite{cacilhas2020controlling}. 
The DW velocity also shows a curvature-dependent oscillatory behaviour in curved nanowires that is distinct from the oscillatory feature above Walker breakdown seen in straight wires \cite{moreno2017oscillatory,thomas2006oscillatory}.

In applications for which DW stochasticity is detrimental, Walker breakdown might need to be minimized and various techniques have been suggested for this. For instance, doping permalloy nanowires with rare-earth elements such as holmium can increase the Gilbert damping parameter and thereby suppress Walker breakdown until much higher magnetic fields \cite{broomhall2017suppression}. Geometric suppression of Walker breakdown in nanowires has also been achieved using a series of cross-shaped wire junctions in a comb-like geometry to interrupt and reset, at regular intervals, the dynamic pathway of DW dynamics that lead to magnetic oscillations \cite{lewis2010fast}. 


We have seen in this and previous sections that DWs in nanowires exhibit rich dynamics which can be controlled via a variety of stimuli including magnetic fields, charge and spin-polarised currents, spin-orbit torques and mechanical strain. DW properties such as nucleation, chirality and position can be controlled by many of these stimuli as well as other factors such defects in the nanowires and by modifying the wire geometry. Furthermore, nanowires show complex behaviour such as stochastic switching of domains involving probabilistic pinning/depinning of DWs. In the next sections, we shall describe applications proposed using DWs in nanowires and explain how, while for some applications these complex effects can be detrimental, for other novel applications, these probabilistic processes can be highly desirable.
\section{Applications of domain walls in patterned nanowires}\label{S4}

Proposed applications of magnetic DWs in patterned nanowires have mostly focused on information technologies, originally just for digital memories and computing, but more recent proposals have included physical realisations of  approaches to unconventional computing. Other uses include magnetic field sensors, various bio-applications using magnetostatic interactions of DWs with magnetic beads, and even externally-addressed cold-atom optics, again using magnetostatic interactions. Here, we review research in each of these areas. Studies which have theoretically or numerically simulated a device will be a (Type:Sim) descriptor while an experimentally demonstrative study will be given a (Type:Expt) descriptor.

\subsection{DW memories}\label{S4-1}

\begin{figure*}
   \centering
       \includegraphics[width=2\columnwidth]{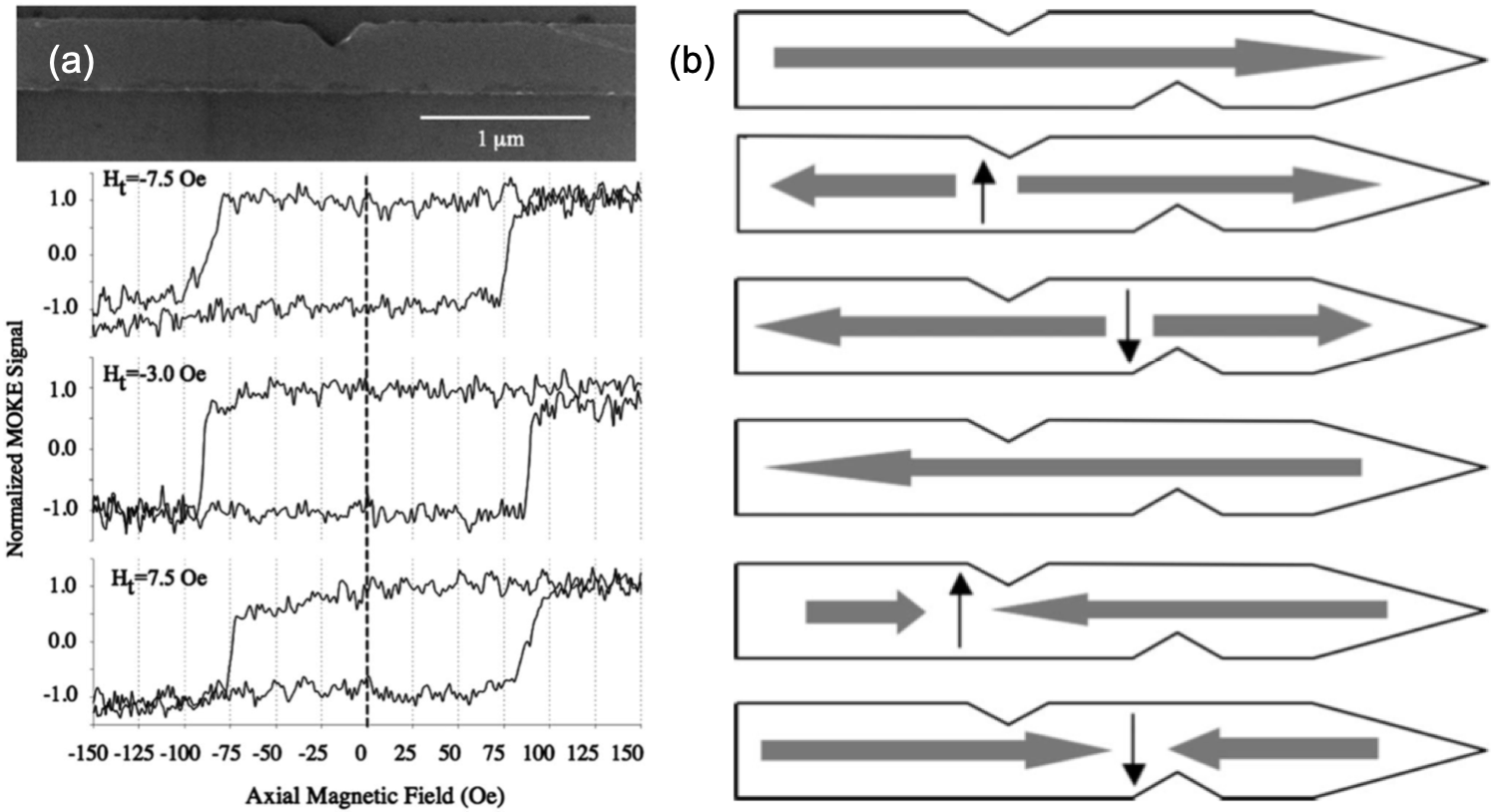}
\caption{(a) An SEM image of a nanowire with a notch with axial field hysteresis loops for different transverse fields to assist depinning. (b) A proposal for a multiple state memory using the direction of magnetisation in the wire and orientation of the DW. Taken from [\onlinecite{atkinson2008controlling}].}\label{fig12}
\end{figure*}

The first magnetic DW-nanowire memory proposal envisaged positioning of a DW between two positions defined by changing lateral dimensions of an IM ferromagnetic wire \cite{mcmichael2000domain} (Type:Sim). The element memory might then be defined either as the magnetisation direction of the central region or the position of the DW, depending on the readout scheme employed. Positional bistability has also been created using induced stress patterns in magnetostrictive materials \cite{dean2011stress}, magnetic fields transverse to the nanowire, particularly when combined with DW chirality \cite{atkinson2008controlling} (refer to Figure \ref{fig12}), and the position of contact wires for current-induced DW motion \cite{numata2007scalable}, which was the basis of the first commercial DW-nanowire MRAM \cite{fukami2009low}. 

Other proposed memories use more continuous propagation of DWs in nanowires. Parkin et al.\cite{parkin2008magnetic,hayashi2008current}  (Type:Expt) proposed a shift register technology known as ‘racetrack memory’ (RM) in 2008 in which digital data are encoded as a series of DWs or oppositely-magnetised domains along an IM nanowire. This approach relies upon using current-induced DW motion in order to ensure that all DWs move in the same direction, thus defining the direction of shift register operation. This approach also offers relatively simple device integration, since only single elements for data writing and readout need be incorporated into an extended wire. Parkin et al \cite{parkin2015memory} (Type:Expt) further proposed different versions of the racetrack with the latest version (v4) making use of synthetic antiferromagnets on heavy metals and exploiting spin-orbit driven effects for driving DWs. These memories promise the advantage of high DW velocity, low power consumption and lesser fringing fields in the device.

Franken et al. \cite{franken2012shift}  (Type:Expt) engineered effective unidirectional field-driven DW motion in an OOPM nanowire ring by patterning a series of gradients in the wire edge profile to make diode structures. These elements prevented back-propagation of DWs, thus allowing a bidirectional out-of-plane magnetic field to result in unidirectional DW motion. The ring structure meant that DWs could continue to circulate controllably for an arbitrary time and fulfil the requirements of a shift register.

Magnetic tunnel junctions (MTJs) \cite{zhu2006magnetic} comprising ferromagnetic layers with free and fixed magnetisation separated by an electrically-insulating layer are also attractive candidates for memory/logic devices. They are particularly useful for reading out the states of devices using phenomena such as tunneling magnetoresistance (TMR) \cite{ikeda2008tunnel}. Lou et al. \cite{lou2008demonstration}  (Type:Expt) have proposed an MRAM device operating at multiple levels implemented using MTJs having a DW in the free layer which are formed due to lithographical imperfections. Subsequently, Raymenants et al. \cite{raymenants2021nanoscale}  (Type:Expt) proposed an improvement to this by introducing a hybrid layer (an additional free layer in an MTJ) to show a three-operation device for DWs that allows addressable writing,  DW transfer, and readout.

In order to increase the storage densities of memories, it is required to minimize the size of the magnetic bit. However, the energy barrier to be overcome for the switching of magnetic domains scales with domain size and thus smaller magnetic bits are more susceptible to thermal effects that can lead to stochastic depinning, which is an existing hurdle towards both dependable switching and long-term information retention in such memories. Another major active research area is the manipulation of DWs for storing information. Magnetic fields control of magnetic states are not attractive for devices due to poor control and high power consumption and so current control of magnetisation is the way forward. However, the current densities required for STT/SOT driven DW motion in nanowires is still relatively high which and their application can lead to significant Joule heating of the device \cite{yamaguchi2005effect}. Furthermore, reading and writing of information requires controlled DW shifting and precise position control (for alignment with read/write heads) and these stringent requirements need further study. These challenges are still to be overcome before the commercialisation of DW based long-term memory storage devices. The interested reader is referred to others reviews \cite{kumar2022domain,kang2019comparative,mittal2016survey} for further information about mechanisms, devices, and materials for DW memories.

\subsection{Boolean computing}\label{S4-2}

\begin{figure*}
   \centering
       \includegraphics[width=2\columnwidth]{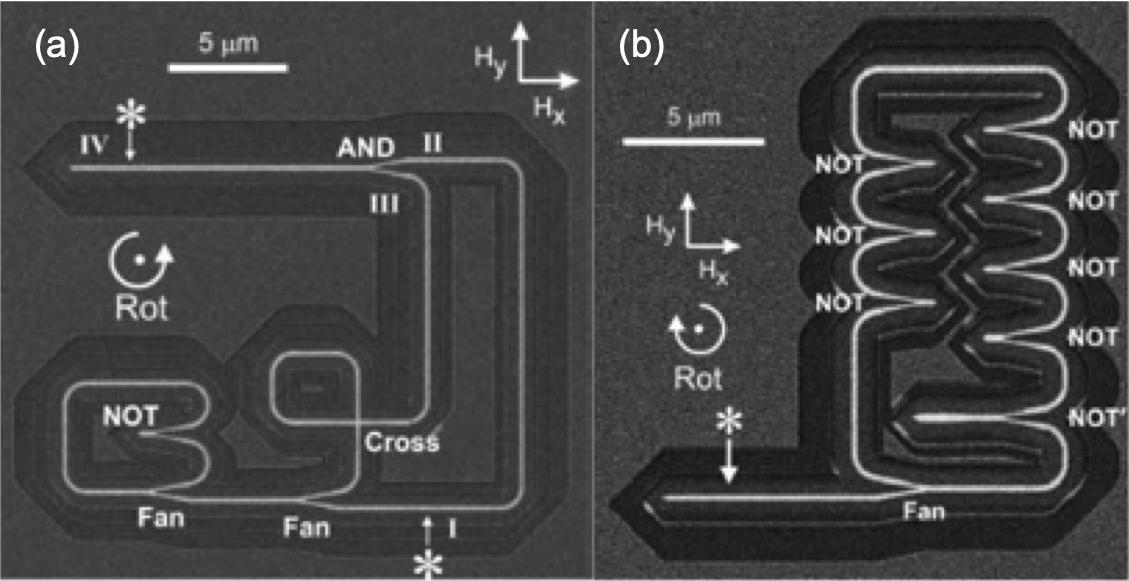}
\caption{Images of (a) a magnetic nanowire circuit showing a combination of different gate logic element implementations mentioned in Table \ref{fig14}. (b) A 5-bit magnetic shift register implemented using NOT gates and a fan-out junction. Taken from [\onlinecite{allwood2005magnetic}].}\label{fig13}
\end{figure*}

\begin{figure*}
   \centering
       \includegraphics[width=2\columnwidth]{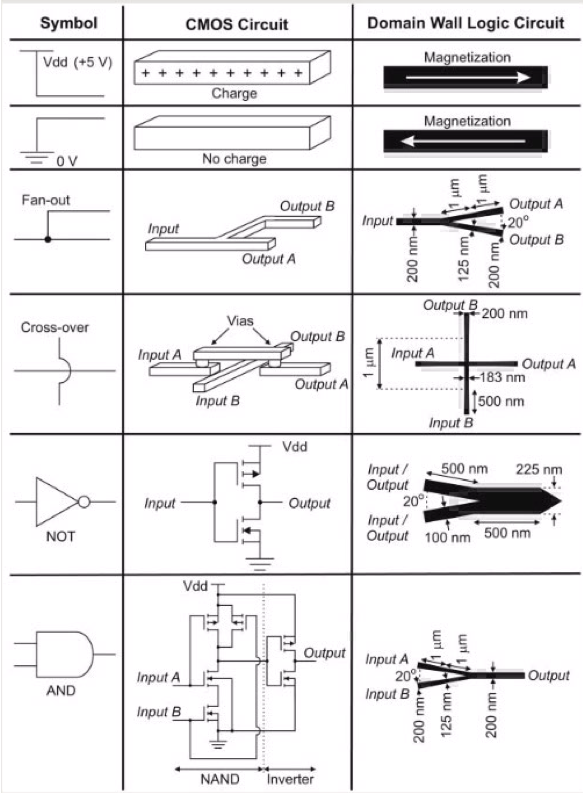} 
\caption{Simple logic elements with their CMOS and magnetic nanowire implementations. Taken from [\onlinecite{allwood2005magnetic}].}\label{fig14}
\end{figure*}

Magnetic-field-driven Boolean computing using DWs in IM nanowires was first demonstrated  approximately 20 years ago by Allwood et al\cite{allwood2005magnetic} (Type:Expt). They used in-plane rotating magnetic fields to drive DWs around 2D nanowire circuits (refer to Figure \ref{fig13}), with wire corners providing a means to limit the extent of DW propagation during any particular phase of field rotation thus allowing the external field to act as a circuit clock. The behaviour of DWs at wire junctions allowed nanowire elements to be created that performed operations usually associated with CMOS logic circuit elements (Figure \ref{fig14}) \cite{allwood2005magnetic}. These included a Boolean NOT-gate, which performed an inversion operation, and an AND/OR-gate, the function of which depended on a dc magnetic field bias direction. They went on to show how these elements could be integrated to realise more complex logic circuits, including using a chain of NOT-gates to create a clocked shift register memory (Figure \ref{fig13}). Allwood et al. \cite{allwood2006magnetic} (Type:Expt) further proposed a shift register using a three-terminal 'Y-shaped' IM nanowire structure with different widths of the nanowires in the structure. While two of the terminals of the device were used for the NOT operation described in\cite{allwood2005magnetic}, the third terminal supplied an extra, non-inverted output and by cascading a series of such structures, they could realised a shift register operation.

\begin{figure*}
   \centering
       \includegraphics[width=2\columnwidth]{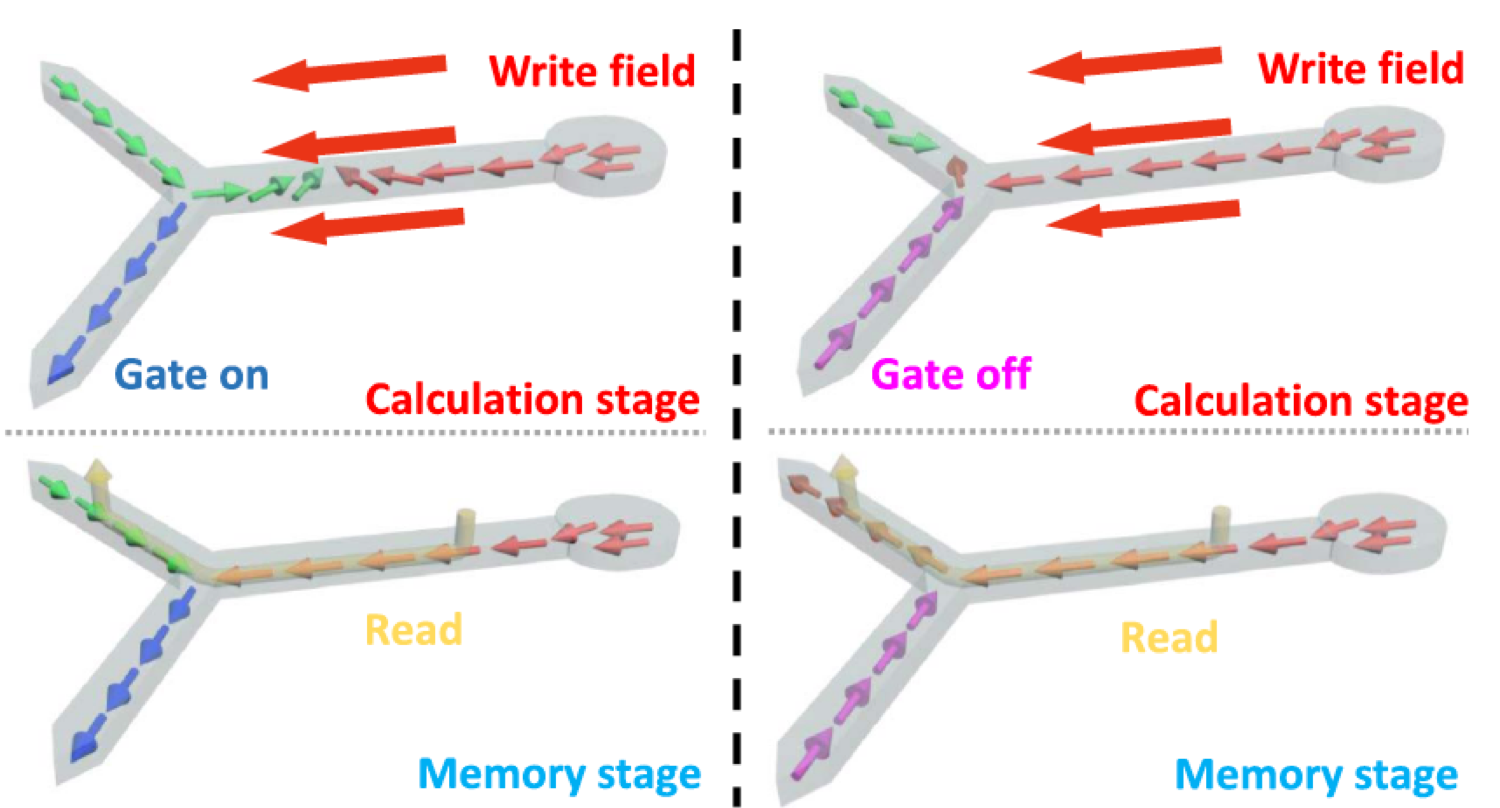}
\caption{The operation of a three-terminal nanowire device with a junction where the magnetic state in one arm ('gate') can be used to control the pinning of a DW at the junction thus control the magnetic states in the other arms, realising a transistor operation. The 'calculation' and 'memory' stage show modes of operation to realise both data processing storage. Taken from [\onlinecite{zheng2020paradigm}].}\label{fig15}
\end{figure*}

Zheng et al. \cite{zheng2020paradigm} (Type:Expt) extended this idea to show that field driven DW motion in a Y-shaped permalloy nanowire device which has a "gate" terminal, the magnetisation of which determines the DW propagation dynamics in the "source" and "drain" arms, can be used to realise a transistor operation. DW depinning at the junction is controlled by the direction of the magnetisation in the gate arm (refer to Figure \ref{fig15}). They proposed in-memory computation using this element where the junction device can be used to store memory states as well as perform logic operations (such as OR and NAND) acknowledging the cross-disciplinary challenges of implementing a spintronic arithmetic logic unit for feasible in-memory processing of data.

The above applications with IM materials used nanowires with dimensions that would result in transverse DWs of a random chirality. Vandermeulen et al. \cite{vandermeulen2015logic} (Type:Sim) proposed a simulated device which used the two chiralities of transverse DWs in an IM nanowire as Boolean logic levels to realise NOT/AND/OR logic gates. Around the same time, Goolaup et al. \cite{goolaup2015transverse} (Type:Expt) demonstrated an IM nanowire device which could be used to invert or rectify the chirality states of transverse DWs. They used magnetic force microscopy to show that the chirality of a T2T transverse DW could be rectified from 'down' to 'up' while in the inverter case, the chiralities of both T2T and H2H DWs could be reversed. The chirality inherent in vortex DWs has also been proposed as the basis of chirality-encoded DW logic \cite{omari2014chirality,omari2019toward}. Omari et al. \cite{omari2019toward} (Type:Expt) used the chirality reversal of DWs that propagate through notches in the wire edges to perform inversion operations. They also showed that two-input wire junctions could be programmed to operate as logic gates (AND/NAND/OR/NOR) by controlling the chirality of the DW in the output arm which in turn was determined by which input arm switched its magnetisation first. 

\begin{figure*}
   \centering
       \includegraphics[width=2\columnwidth]{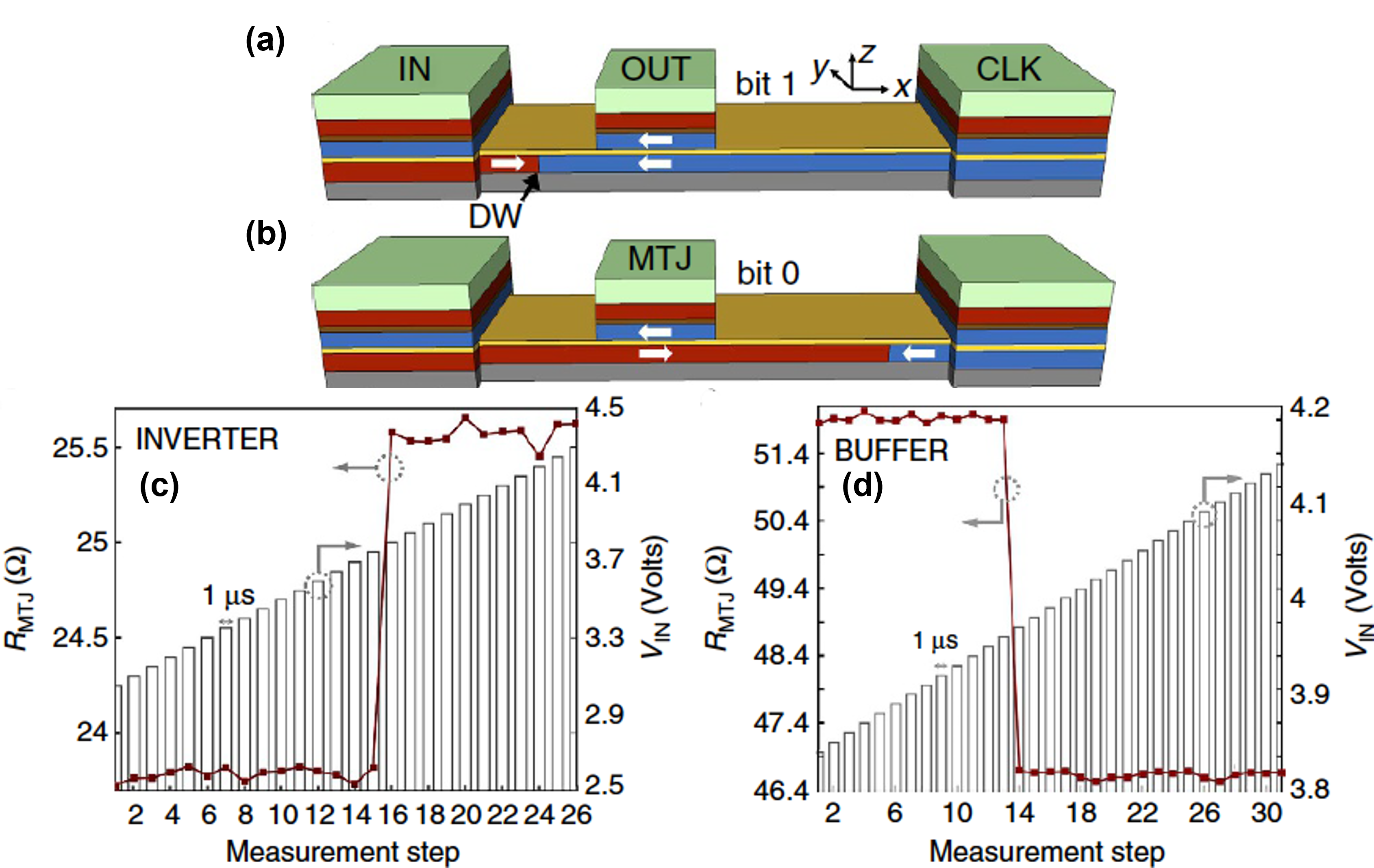}
\caption{(a) and (b) A DW in a nanowire read out using the resistance state of an MTJ on top of the nanowire. The position of the DW denotes a logic '0' or '1'. (c) and (d) The device showing an inverter and buffer operation on applying clock pulses. Taken from [\onlinecite{currivan2016logic}].}\label{fig17}
\end{figure*}

The above proposals and demonstrations were with magnetic fields being the agents to manipulate DWs. We now consider electric current driven proposals for Boolean logic devices.  Incorvia et al. \cite{currivan2016logic} (Type:Expt) demonstrated a three-terminal DW-based MTJ device for various logic operations including inverter and buffer operations. A memory cell device was modified to be used for logic purposes with input information encoded into the position of a single DW (as a logic '0' or '1' as shown in Figure \ref{fig17} (a) and (b)) in a nanowire using STT and the information state was read out using TMR. They demonstrated an inverter and a buffer operation (refer to Figure \ref{fig17} (c) and (d)) using the device and subsequently realised a three inverter network by arranging three devices in series.
Manfrini et al. \cite{manfrini2018interconnected} (Type:Expt) used a four-pillar device to propagate DWs over long distances in the common free layer of four MTJs using STT with the device state read out using TMR of the output pillar. They then used these device states to simulate the implementation of a spin torque majority gate which is a candidate for an efficient spin-logic device \cite{nikonov2011proposal}. 

\begin{figure*}
   \centering
       \includegraphics[width=2\columnwidth]{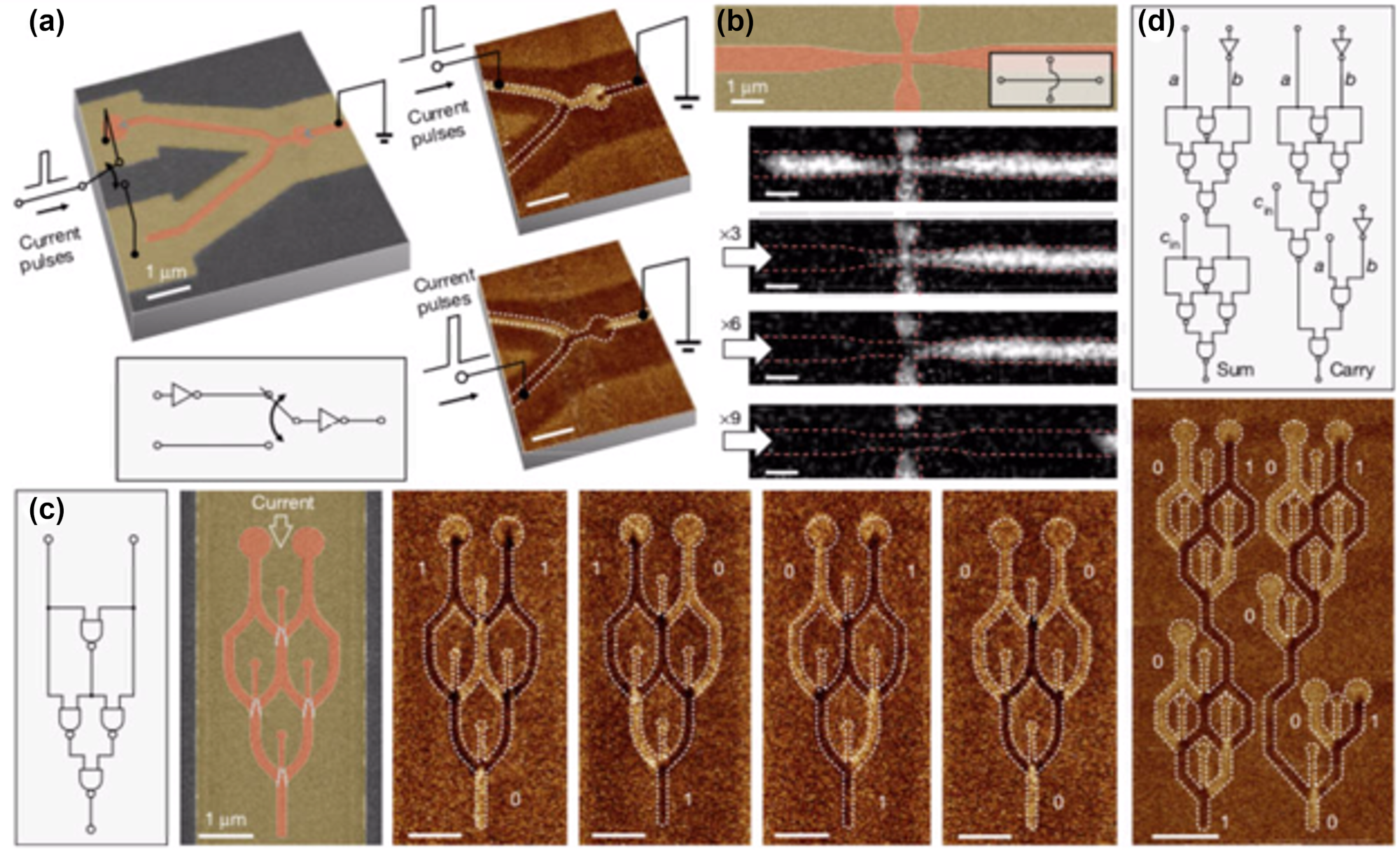}
\caption{(a) A Y junction device which is demonstrated to be a logic circuit.(b) MOKE images showing operation of the junction device when driven using current pulses. (c) A XOR gate demonstrated using NAND gates and (d) a full adder circuit with magnetic force microscopy images showing different states in the gate operations. Taken from [\onlinecite{luo2020current}].}\label{fig16}
\end{figure*}

While electric current driven control of DWs in nanowires have many advantages over field driven control (refer to Section \ref{S2-3}), the challenges associated with electric field control of DWs (such as high current densities required) has limited device proposals. While current-driven DW-based NOT-gates were demonstrated in 2008 using Invar elements \cite{xu2008all}, it was the use of SOT driven current to control DWs that has led to progressing the state of the art involving proposals of current driven DW based logic devices. Baek et al. \cite{baek2018complementary} (Type:Expt) have shown that a combination of electrical-field-controlled SOT switching and voltage-controlled magnetic anisotropy switching of the magnetisation in $\rm{Ta/CoFeB/MgO/AlO}_{\rm{x}}$ can be used to realise a spintronic logic device. They claimed that compared to a CMOS implementation, a half-adder implemented using their system would be an order of magnitude smaller, have lower energy consumption, and offer the flexibility of dynamic reconfigurability. Subsequently, Luo et al. \cite{luo2020current} (Type:Expt) were able to demonstrate all-electrical control of DWs in OOPM nanowires to perform all necessary operations for a full logic architecture (Figure \ref{fig16}). This included development of three-input-one-output wire junctions to create majority gates with a logical AND/OR function governed by the magnetisation direction of a central control wire, and half-and full-adder circuits made of cascaded magnetic nanowire NAND-gates (Figure \ref{fig16} (c)-(d)). This approach offers the potential of highly efficient and high-speed processing based on DWs in magnetic nanowires.

\begin{figure*}
   \centering
       \includegraphics[width=2\columnwidth]{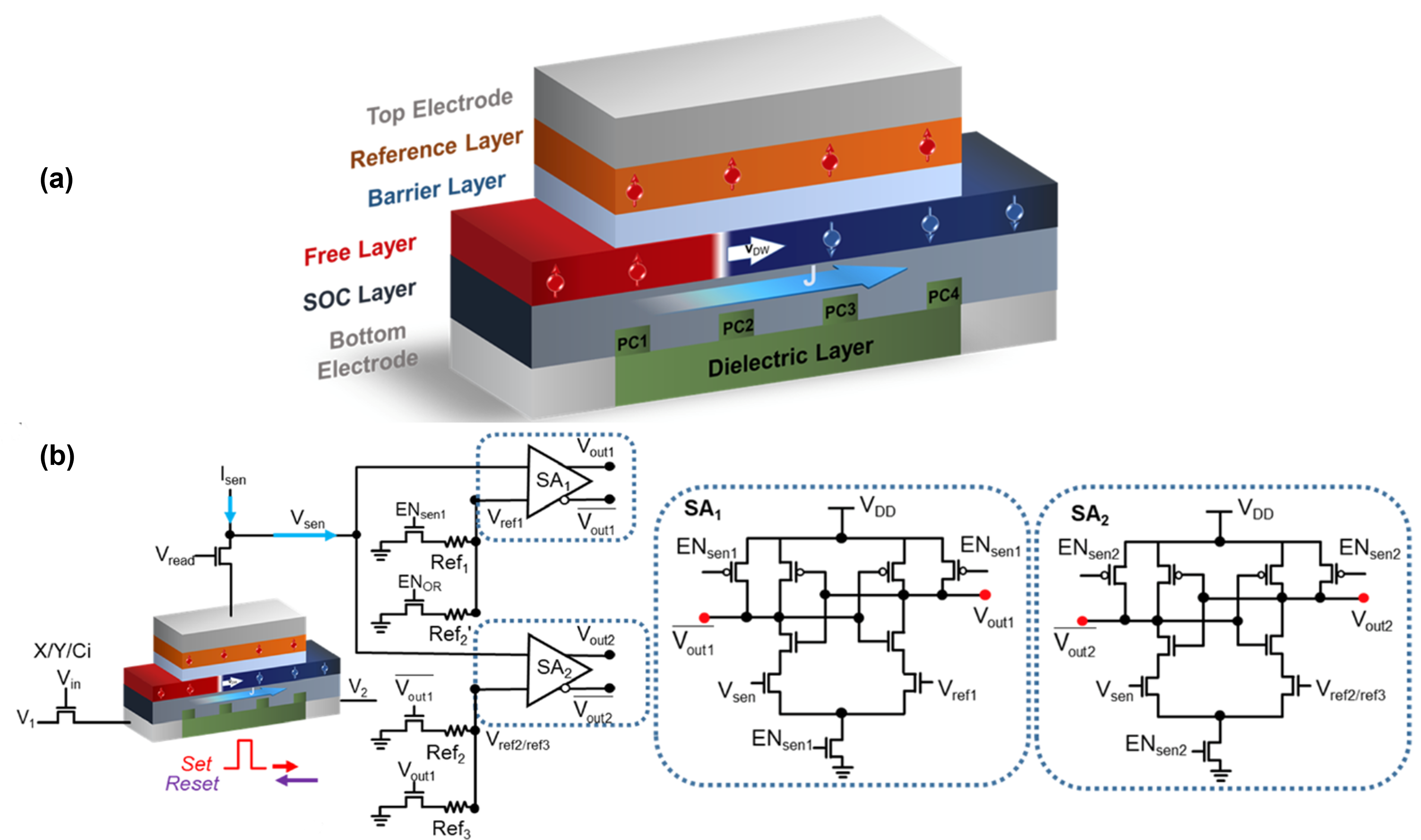}
\caption{(a) The proposed SOT/DW/MTJ device for in-memory computation. (b) The proposed hybrid device to make use of multiple DW states for implementing logic operations. Taken from [\onlinecite{lin2022implementation}].}\label{fig18}
\end{figure*}

We now shift our attention to OOPM materials based proposals and demonstrations. Alamdar et al. \cite{alamdar2021domain} (Type:Expt) drove OOPM Ta/CoFeB/MgO DW-MTJ heterojunction devices using SOT switching to realise two device inverter circuits by controlling the DW position in the CoFeB layer, which in turn determined the output resistance state of the MTJ. They optimised the PMA and lithography process to obtain a high TMR ratio and suitable average resistance-area product of the device. Lin et al. \cite{lin2022implementation} (Type:Expt) have proposed a multistate SOT-DW-MTJ device (refer to Figure \ref{fig18} (a)) for in-memory computation applications. Apart from the three typical layer stack of an MTJ (they considered CoFeB/MgO/CoFeB), the device has a W layer on each side of the stack to control the position of the DW in the free CoFeB layer using SOT generated by passing current pulses through the W layer. Along with MOKE imaging of the DW states in similar devices, they simulated the operation of buffer, inverter and typical logic gates on single and dual inputs to the device. They also used micromagnetic and circuit level simulations to show the operation of a full adder using their device and expect the read/write latency to be as short as $1.25/0.22\,\rm{ns}$, and average writing energy of $8.4\,\rm{fJ/bit}$ and a power consumption of $26.25\,\mu\rm{W}$ which they claim is significantly lower than alternative implementations of devices for similar applications. This is promising as according to previous estimates by Xiao et al. \cite{xiao2019energy} the power consumption of a 32 bit STT/SOT driven DW-MTJ adder circuit is comparable to its CMOS counterpart.

\begin{figure*}
   \centering
       \includegraphics[width=2\columnwidth]{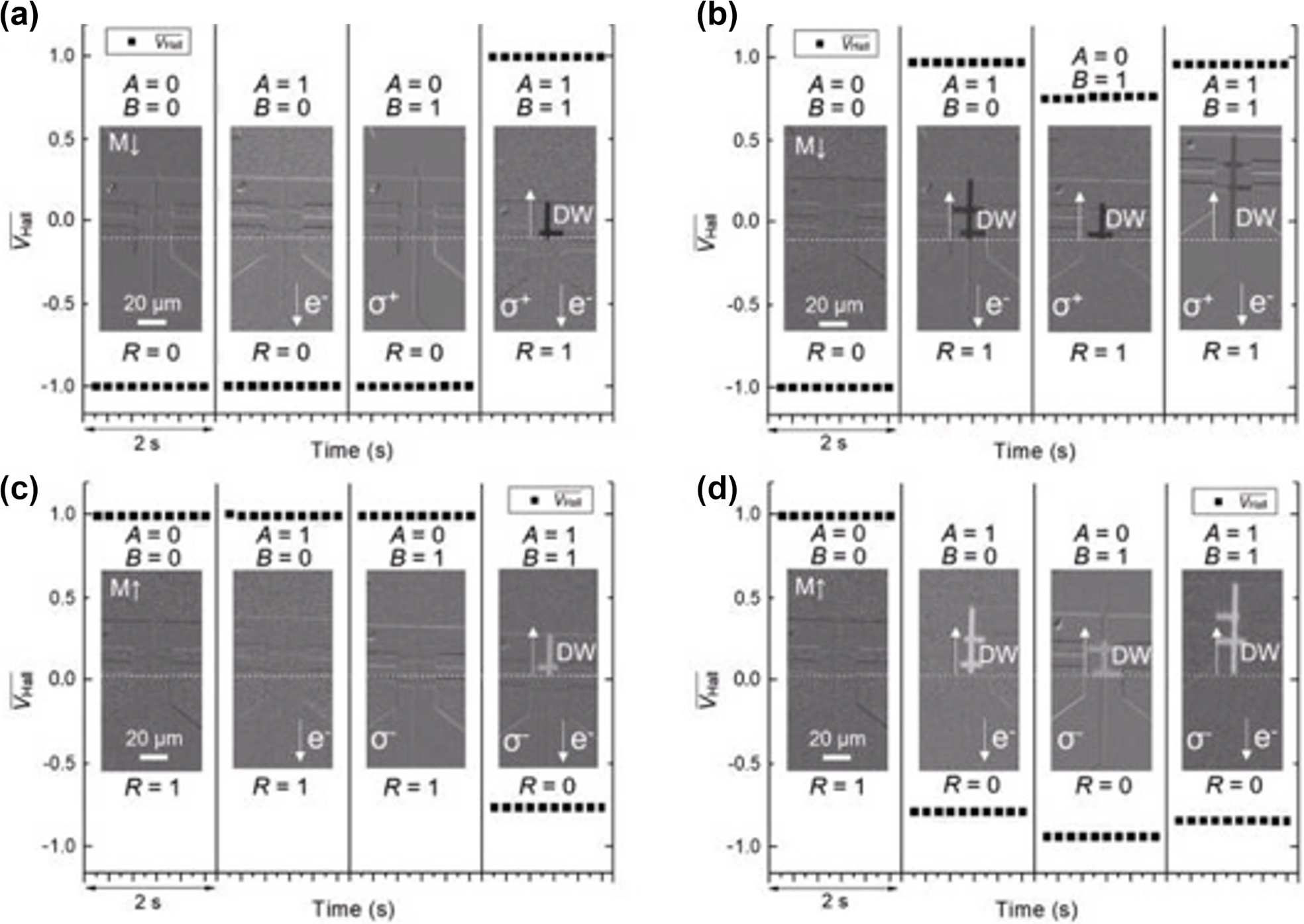}
\caption{(a) Opto-electronically driven DW motion used for realising: (a) AND, (b) OR, (c) NAND, and (d) NOR logic gates. Inputs $A$ and $B$ correspond to current and optical pulses used to drive DWs and $R$ denotes the resistance state of the wire which serves as the logic gate output. Taken from [\onlinecite{zhang2020optoelectronic}].}\label{fig19}
\end{figure*}

Finally, in proposals involving different DW control stimuli, Zhang et al. \cite{zhang2020optoelectronic} (Type:Expt) have supplemented the use of current-induced DW motion with ultrafast laser pulses of controlled helicity in a nanowire device. The current pulses had to be timed with the application of optical pulses and allowed a range of logic operations to be demonstrated at wire junctions, as shown in Figure \ref{fig19}. Due to requiring lower current densities to drive DWs. the energy required for this implementation is also lower than tradition current driven DW motion. 

The SOT driven control offer the possibility of device operation at reasonable current densities and this creates industrial potential in these devices. Moreover, as most of the demonstrations in which individual DW based logic devices are integrated use CMOS compatible interconnects, this should be amenable for industrial development. There are a significant number of challenges that need to be overcome before commercialisation. The individual logic elements (like a Boolean logic gate) need to scaled down in size for incorporating advanced logic function implementations in DW based Boolean logic devices. Moreover, more complex logic and networks need to be demonstrated and studied for more complex computation and these need to be tested for repeatability and reliability. They need to be engineering for precise control of DW manipulation at high speeds of operation. The stability of DWs becomes are a significant issue here as well and in the next section we describe device proposals which aims to use this instability of DWs for computing operations.   

\subsection{Non-Boolean/Neuromorphic/Unconventional computing}\label{S4-3}

Neuromorphic computing (NC) describes various forms of ‘brain-inspired’ computing \cite{krittanawong2017artificial,gaba2013stochastic,raza2015review,christensen20222022,hoffmann2022quantum} and is seen as offering exciting routes to reducing the power consumption involved in implementing artificial intelligence algorithms currently implemented using CMOS-based von Neumann architectures \cite{gonzalez1996energy,bhatti2017spintronics}.
NC involves replacing part or all of a conventional computer with bespoke elements that enable computation. One common such element is a neural network, in which a neuron acts as a processing element that transforms multiple inputs into a processed output. A synapse is a memory element that has a multi-weighted transmission response and conducts neuron outputs, with weights generally trained (using a training algorithm) to solve a particular task. The response of the neuron to a stimulus, called the ‘activation function’, should have a required degree of nonlinearity to achieve efficient separation of inputs in classification tasks. Neurons may also have more advanced functions including 1) an 'integrate-and-fire' response which is outputting an input signal when its integrated value crosses a threshold, 2) a 'leaking' property by virtue of which the energy stored in the neuron is dissipated over time and 3) 'lateral inhibition', by which a firing neuron disables other neurons in the same layer from firing. These properties will be useful for understanding spintronic proposals/implementations of NC.

Spintronic implementations of NC \cite{grollier2020neuromorphic} have been demonstrated in recent years using spin-torque nano-oscillators \cite{torrejon2017neuromorphic}, superparamagnetic tunnel junctions \cite{vodenicarevic2017low}, spin waves in a ring resonator \cite{watt2020reservoir}, and artificial spin ices (ASIs) \cite{gartside2022reconfigurable}. They have also been proposed using interactions in spin-Hall nano-oscillators \cite{zahedinejad2020two}, ensembles of superparamagnetic particles \cite{welbourne2021voltage}, skyrmions \cite{prychynenko2018magnetic,song2020skyrmion,pinna2020reservoir}, dipolar-coupled nanomagnets \cite{bhanja2016non}, magnetic states of ASIs \cite{jensen2018computation} and spin waves in a thin film \cite{nakane2018reservoir,dale2021computing} and in coupled nanomagnets on a thin film \cite{papp2021characterization, papp2021nanoscale}. Here, we shall focus on approaches that use magnetic DWs in nanowires to perform NC. Table \ref{tab1} shows a comparative analysis of all these approaches with prospective advantages and challenges.

\begin{table*}
\begin{tabularx}{\textwidth}{ |X|X|X|X| } 
 \hline
 \textbf{Type of spintronic element} & \textbf{Computation performed/proposed} & \textbf{Advantages} & \textbf{Challenges} \\ 
 \hline\hline
 Spin torque nano-oscillators and superparamagnetic tunnel junctions & Magnetic synapse \cite{lequeux2016magnetic}, Spoken digit recognition and waveform classification \cite{torrejon2017neuromorphic}, vowel recognition \cite{romera2018vowel}; generation of cursive letters \cite{mizrahi2018neural}, stochastic computing \cite{daniels2020energy}, random number generation \cite{vodenicarevic2017low} (Type:Expt) & Small size, low power consumption, CMOS compatibility, downsizing to potentially atomic dimensions, switching at low current densities (due to aid of thermal perturbations) & Complicated fabrication of STNOs, Low resistance difference between ON and OFF states of MTJ make reading large arrays non-feasible, high phase noise, trade-off between energy cost and correlation of random number generators \\ 
 \hline
 Spin waves in films & RC properties \cite{watt2020reservoir} (Type:Expt), time-multiplexed multi-dimensional input for reservoir computing  \cite{watt2021implementing} (Type:Expt), signal estimation \cite{nakane2018reservoir} (Type:Sim), reservoir metrics and nonlinear autoregressive moving average \cite{dale2021computing} (Type:Sim), frequency detection \cite{papp2021nanoscale} (Type:Sim) and evaluation of task agnostic RC metrics \cite{papp2021characterization} (Type:Sim) & Low power, short wavelengths, nonlinear and anisotropic interactions & High quality growth of films for low damping, may show best performance at low temperatures \\
\hline
Artificial spin ices & Time series prediction \cite{gartside2022reconfigurable} (Type:Expt), multi-level and deep RC networks \cite{stenning2022adaptive} (Type:Expt) and pattern classification \cite{jensen2018computation} (Type:Sim) & Large number of strongly interacting elements with minimal need of interconnects, complex and tunable dynamics & High resolution fabrication required (elements are typically $100-200\,\rm{nm}$), experimental demonstration challenging due to weak signals\\
\hline
Skyrmions & skyrmion neurons \cite{li2017magnetic, he2017tunable} (Type:Sim), skyrmion synapses \cite{huang2017magnetic} (Type:Sim), synaptic pattern recognition \cite{song2020skyrmion} (Type:Expt) and RC Pattern recognition \cite{pinna2020reservoir} & Complex dynamics and strong interactions in close proximity, relatively low current densities for positioning & Thermal noise might swamp AMR signals from devices, complex fabrication requirements, stability regions are difficult to achieve on a device. \\
\hline
Spin Hall nano-oscillators & Synchronisation maps useful for neuromorphic computation \cite{zahedinejad2020two} (Type:Expt), addition of two neuron outputs and multiplication of neuron output by synaptic weights \cite{markovic2022easy} (Type:Sim) & Easier fabrication than STNOs, mutual synchronisation possible17 & Non-volatile storage of oscillator state a challenge, nonlinear and complex interactions between oscillators to be better understood \\
\hline
Superparamagnetic particle ensembles & Spoken digit recognition and time series prediction \cite{welbourne2021voltage} (Type:Sim) & Low power consumption, Large range of response timescales tuned by voltage & Yet to be fabricated, response reproducibility for a device maybe challenging \\
\hline
Dipolar coupled nanomagnets & Perceptual organisation in computer vision \cite{bhanja2016non} (Type:Expt) and waveform classification & Small scale elements and replacement with MTJs can give a technological boost & Readout done only using imaging techniques21 and electrical readout is envisaged  \\
\hline
\textbf{Domain wall based devices} & DW-based neuron \cite{leonard2023stochastic} (Type:Expt), DW-based synapse \cite{siddiqui2019magnetic, leonard2022shape} (Type:Expt), RC signal reconstruction, spoken digit recognition and reservoir computing metrics (experimental \cite{vidamour2022reservoir} and simulation \cite{vidamour2022quantifying, dawidek2021dynamically}) & Relatively simple fabrication, multiple methods of controlling DW states, easy readout using electrical means, complex dynamical phenomena & Current and SOT control and CMOS compatibility are yet to be investigated for unconventional computing   \\
\hline
\end{tabularx}
\caption{Unconventional computing implementations using magnetic systems with potential advantages and challenges.}\label{tab1}
\end{table*}

\begin{figure*}
   \centering
       \includegraphics[width=2\columnwidth]{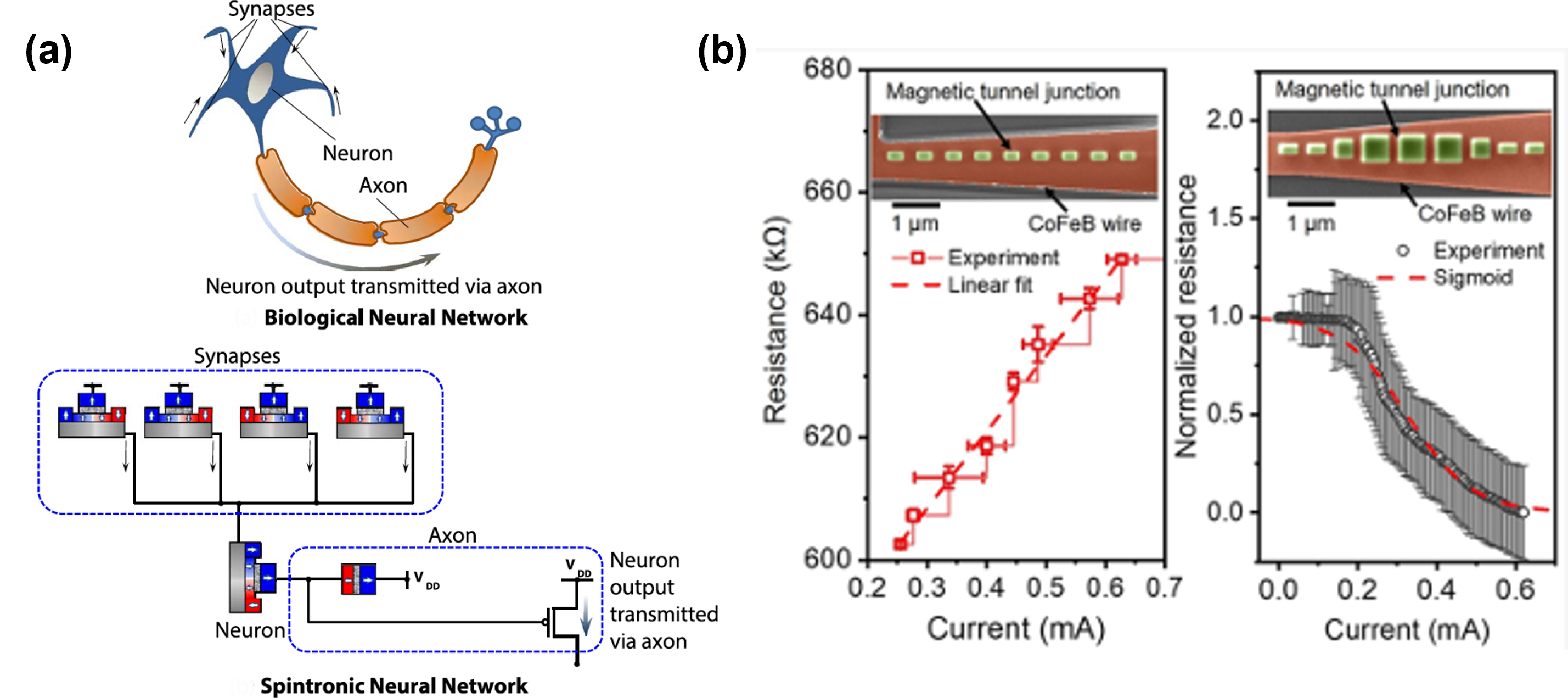}
\caption{(a) A combination of a biological synapse and neuron with its proposed spintronic equivalent. Taken from [\onlinecite{sengupta2016proposal}]. (b) Linear and nonlinear activation functions realised using an MTJ based device. Taken from [\onlinecite{siddiqui2019magnetic}]}.\label{fig20}
\end{figure*}

A DW-based spintronic implementation of a neuron and a synapse was initially proposed in 2016 \cite{sengupta2016proposal} (Type:Sim) using MTJs (refer to Figure \ref{fig20} (a)) with OOPM materials used for the free and fixed layers with the free layer having a DW. The neuron implementation was suggested to be achieved by moving the position of the DW in the free layer using SOT generated by current pulses. Synaptic behaviour was envisaged via a multi-step conductivity response of the MTJ obtained by moving the DW in the free layer and between artificial pinning sites. Subsequently, Hassan et al. \cite{hassan2018magnetic} (Type:Sim) proposed a neuron with lateral inhibition by extending the DW-based MTJ proposed by Incorvia et al.\cite{currivan2016logic} to include an underlying hard IM ferromagnetic track. The integrate (by passing a current through the DW track) and firing occurred when the DW reached an MTJ at the end of its track causing the MTJ's resistance state to change. Leaking action was obtained by the dipolar field of the ferromagnetic track on the DW in the MTJ free layer, causing the DW to return to its initial state after the application of the current pulse. They went on to propose the arrangement of multiple devices in an array and could obtain lateral inhibition due to the suppression of DW movement in one track by dipolar coupling with its neighbouring track. Modification of the spacing between tracks and the current passing through them could control the degree of lateral inhibition and the array response was also used to categorise more than 100 handwritten digits with $94\%$ accuracy. 

Further recent proposals of NN functions with DW motion include utilising magnetocrystalline anisotropy gradients \cite{brigner2019graded} and by using width modulated wires \cite{brigner2019shape}, both of which modify the energy landscapes of DWs for propagation. The leaking action is provided by a resorting force which is induced by the modified energy landscape on a displaced DW. Brigner et al. \cite{brigner2019graded} (Type:Sim) suggest that by introducing anisotropy gradients in a 3-terminal MTJ device, DWs move from regions of higher to lower anisotropy without external stimuli which leads to a form of leaking action. Upon passing a current through the DW track, the DW movement causes the integration of the current response. Similar to the above, field and current free DW movement due to the width gradient in a nanowire \cite{brigner2019shape} (Type:Sim) causes the simultaneous implementation of the leaky and integrate actions required for a neuron. Cui et al. \cite{cui2020maximized} (Type:Sim) simulated current and field-driven DW motion in a pair of adjacent MTJ-DW devices and obtained lateral inhibition by tuning the magnetostatic interaction between the devices at the Walker breakdown field. Wang et al. \cite{wang2022stochastic} (Type:Sim) have used micromagnetic simulations of DWs in an OOPM nanowire in which the DW nucleation is stochastic using STT (due to thermal effects) and DW dynamics are deterministic using SOT. This led to both stochastic modifications of synaptic weights and a multilevel output synaptic response. They could simulate machine learning of breast cancer data with an accuracy of $95.7\%$ accuracy and estimated an energy consumption of less than $2\,\rm{fJ}$ for device based on their proposal.

Experimental realisations of NC based elements using DW-based MTJ devices have concentrated on synaptic behaviour and Siddique et al. \cite{siddiqui2019magnetic} (Type:Expt) showed both linear and nonlinear activation functions by using patterned MTJ elements on an underlying DW-carrying $\rm{CoFeB}$ nanowire (which was the MTJ free layer). DW positions were changed by SOT induced by currents passed through the free layer and read out via the resistance states of output MTJs. This approach allows the precise form of activation function to be specified by the relative size of successive MTJ regions along the DW nanowire conduit, for example, using successively larger and then smaller MTJs creates a sigmoidal response (refer to Figure \ref{fig20} (b)). Furthermore, it is relatively fast, being operated with $8\,\rm{ns}$ current pulses, and energy efficient, since each pulse consumed $<16\,\rm{pJ}$. 
Leonard et al. \cite{leonard2022shape} (Type:Expt) have demonstrated a DW-based synapse read out using a single MTJ. The device consisted of an MTJ stack with an OOPM $\rm{CoFeB}$ free layer and an underlying Ta layer for writing magnetic states using SOT. The $\rm{CoFeB}$ layer had a DW track patterned notches so as to control DW nucleation and pinning. The MTJ resistance was read out using TMR via electrical contacts and they considered trapezoidal and rectangular shaped DW tracks for different functionalities. Since the trapezoidal device had a varying tack width and the threshold  voltage  to  depin the DW is inversely proportional to the track width, it showed deterministically switched resistance states. Also, while it had 9 intermediate notches, the authors were able to obtain only 4 resistance states from it and they attributed this to inconsistencies in the fabrication process. The rectangular device, on the other hand, had constant width and notch configurations and showed probabilistically switched resistance states. The authors subsequently simulated machine learning of a Fashion-MNIST dataset (using the trapezoidal geometry) and of an  CIFAR-100  image  recognition (using the rectangular geometry). Importantly, the authors also compare the performance of their DW-based synapse with other proposed synapse implementations in terms of write energy, update duration and write noise. Leonard et al. \cite{leonard2023stochastic} (Type:Expt) have also shown DW-based device with MTJ readout that used a similar stack configuration and SOT driven stimulus as used in \cite{leonard2022shape}. The device had a voltage dependent activation function and they used this response to implement machine learning using a noisy version of the Fashion-MNIST dataset and claimed that such devices could be used to implement robust networks for NC. The interested reader can refer to the recent review by Hu et al. \cite{hu2023magnetic} for more details on such devices.

Kumar et al. \cite{kumar2023ultralow} have proposed a DW based synapse using a meandering OOPM nanowire configuration. The magnetic states were written using a combination of an OOP magnetic field and SOT via an underlying Tungsten layer. This layer was grown using a novel procedure involving sub layers grown at different pressures and this reduced the roughness of the Tungsten on contact with the magnetic layer and aided spin current propagation. This led to low SOT current densities being required $\sim10^{6}\,\rm{A}/\rm{m}^{2}$ and a low power consumption of $0.4\,\rm{fJ}$ to move a DW by $\sim19\,\mu\rm{m}$. The authors varied the distance between neighbouring segments in the meandering nanowire to realise synapses of varying degree of stochasticity.

The stochastic dynamics of DWs at notches in IM nanowires used have also been used to to demonstrate functional binary stochastic synapses by Ellis et al. \cite{ellis2023machine} (Type:Expt). They used magnetic field driven nucleation and Oersted fields of current pulses for DW dynamics in an IM permalloy nanowire with an artificial notch to realise the sigmoid-like passing probability of a DW using MOKE. They went on to simulate and demonstrate machine learning of handwritten digit recognition in the device using a gradient learning rule that adjusted synaptic stochasticity and energy throughput depending on the number of measurements.

All these explorations indicate that there is promise in realising NC devices and circuits using DW-based devices. There is now a need to scale the elements of this implementations to realise more complex circuits and realise further novel functionalities. The realisation of neurons and the other proposed foundational NC elements are necessary for this purpose. Furthermore, the possible fabrication and behavioural inconsistencies possible with DW tracks and nanostructures need to be resolved for technological implementations.

\begin{figure*}
   \centering
       \includegraphics[width=2\columnwidth]{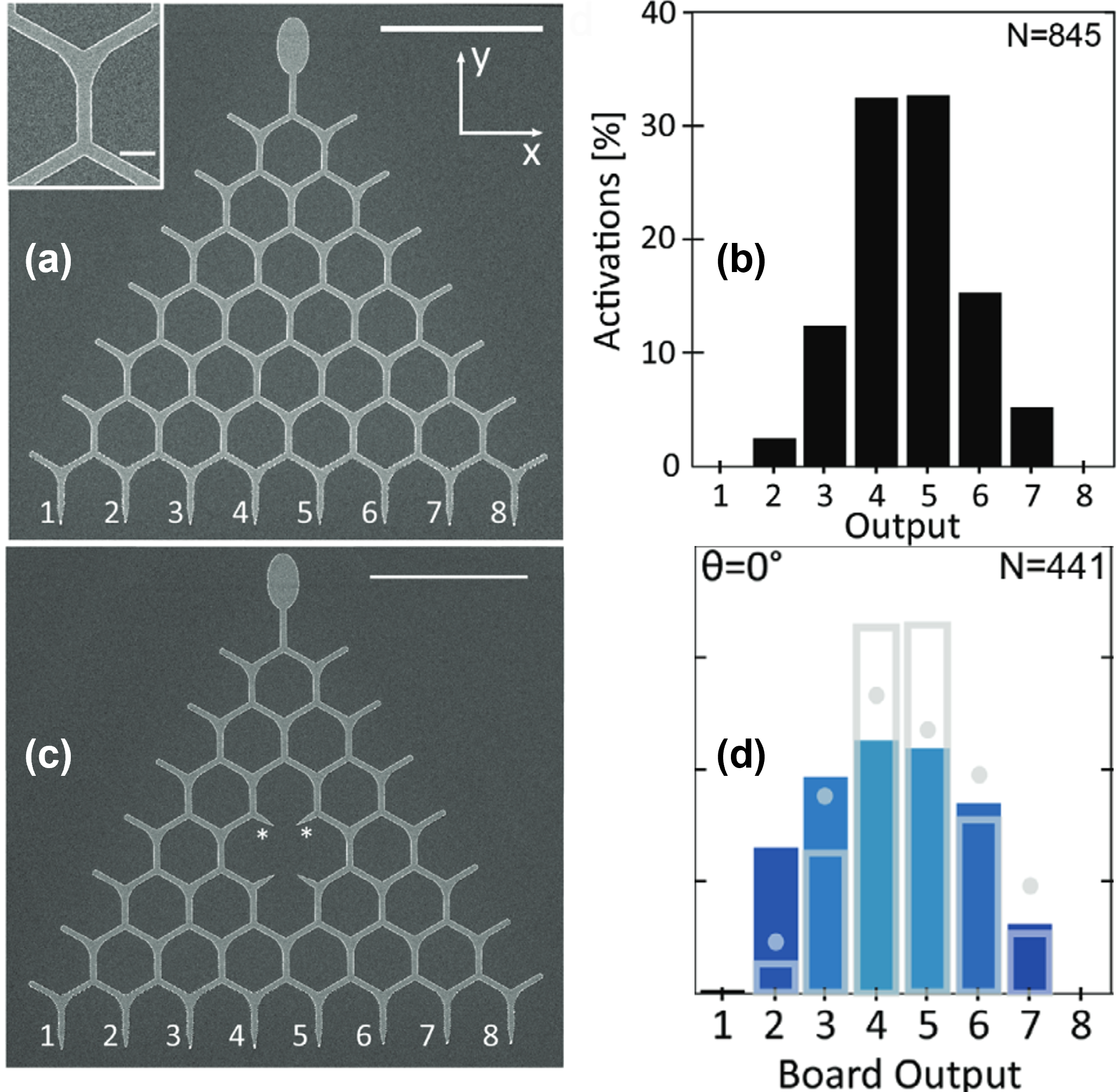}
\caption{SEM micrographs of a magnetic nanowire Galton board which are (a) complete and (c) with the central element missing. The distribution of outputs of (b) the complete board which approached a binomial distribution and (d) the board with an element missing. Here the distribution is flattened indicating tuning of the board response. Taken from [\onlinecite{sanz2021tunable}]}\label{fig21}
\end{figure*}

Apart from proposals for neural network elements, other forms of unconventional computing have also been explored. One such paradigm is stochastic computing where complex computations can be realised using streams of random bits \cite{alaghi2013survey}. In this context, the inherent uncertainties of DW dynamics can also be useful for stochastic computing paradigms \cite{alaghi2017promise}. Hernandez et al. \cite{sanz2021tunable} (Type:Expt) have used the stochastic pinning and propagation of DWs at nanowire junctions DWs to realise a Galton board, an archetypal experiment in statistics. The original Galton board consisted of balls moving down a slanted surface encountering a grid of obstacles which led to a Gaussian diffusion random walk of the balls. In [\onlinecite{sanz2021tunable}], a magnetic field was used drive a single DW in a branched network of IM magnetic nanowires to recreate a Galton board (Figure \ref{fig21} (a)). The distribution of DWs across the eight output wires approximated a normal function (Figure \ref{fig21} (b)), although the final position of successive DWs was highly uncorrelated. The authors passed the binarised output sequence through the NIST Statistical Test Suite for Random and Pseudorandom Number Generators and it passed all 13 tests indicating a high degree of uncorrelation. Interestingly, removing the central wires in a separate structure resulted in a flattened DW distribution across output wires (Figure \ref{fig21} (c) and (d)). This highlights how this simple approach could be adapted to different tasks. One might envisage adaptive, addressable tuning of DWs at wire junctions offering training functionality in reconfigurable magnetic nanowire Galton board networks.

Another random number generator was demonstrated by Narasimman et al. \cite{narasimman2020126} (Type:Expt) who reported a $65\,\rm{nm}$ CMOS readout circuit for an OOPM Hall cross in which the magnetic texture took random orientations due to the thermal and demagnetization effects. They changed the magnetic textures using pulsed currents and detected these textures using the anomalous Hall effect by a modulation and amplification scheme. Their device consumed a power of $126\,\mu\rm{W}$ when driven using a $1.2\,\rm{V}$ supply and could function with voltage drifts of $440\,\mu\rm{V}$.

\begin{figure*}
   \centering
       \includegraphics[width=2\columnwidth]{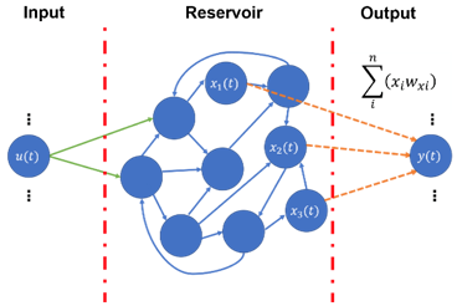}
\caption{(a) A typical implementation of RC, highlighting the layered architecture: an unchanged reservoir layer with individual neurons has time-varying signals fed into it via weighted connections. A trainable output layer then provides a weighted sum of nodal states from the reservoir. Taken from [\onlinecite{vidamour2022quantifying}]}.\label{fig22}
\end{figure*}

The  final unconventional computing paradigm that we shall consider is that of reservoir computing (RC), which is a bio-inspired computational paradigm like NC and has emerged as a prime contender for processing time-varying signals \cite{jaeger2001echo,tanaka2019recent}. A reservoir computer typically consists of an input layer, a reservoir layer and an output layer (refer to Figure \ref{fig22}). A conventional reservoir consists of a recurrent neural network with fixed internal weights. The recurrence in the network, the neuron interactions, and the behaviour of individual neurons themselves combine to provide a time-dependent nonlinear reservoir response to input, which makes RC well-suited for analysing time-varying data. However, unlike traditional recurrent neural networks, only the output layer has to be trained, which makes training very efficient. 

The fixed internal weights of the reservoir implies that it can be replaced using any dynamical system and this has led to many \emph{in-materio} RC schemes, which involve using a physical system’s complex response to a stimulus as the reservoir transformation. Ideally, the physical reservoir should offer a means to represent data, nonlinear sensitivity to input, a memory of previous input that recedes over time (‘fading memory’), and have readable output. In practice, it should also be scalable and exhibit reproducible device dynamics. Magnetic systems are attractive candidates for reservoirs due to their inherent hysteretic and nonlinear response to stimulus, e.g. magnetic field or SOT. There are also a host of methods for representing, inputting, and reading data, as we have described here for DW-based nanowire systems.  This promise is reflected in the diversity of studies that have proposed magnetic devices for RC, including using skyrmions \cite{pinna2020reservoir}, magnetic states in ASIs \cite{jensen2018computation}, and superparamagnetic particles \cite{welbourne2021voltage}, and demonstrations using STNOs \cite{torrejon2017neuromorphic} and spin waves in ASIs \cite{gartside2022reconfigurable}. The interested reader can refer to the review by Allwood et al. \cite{allwood2023perspective} for further details about RC using nanomagnetic devices.

\begin{figure*}
   \centering
       \includegraphics[width=2\columnwidth]{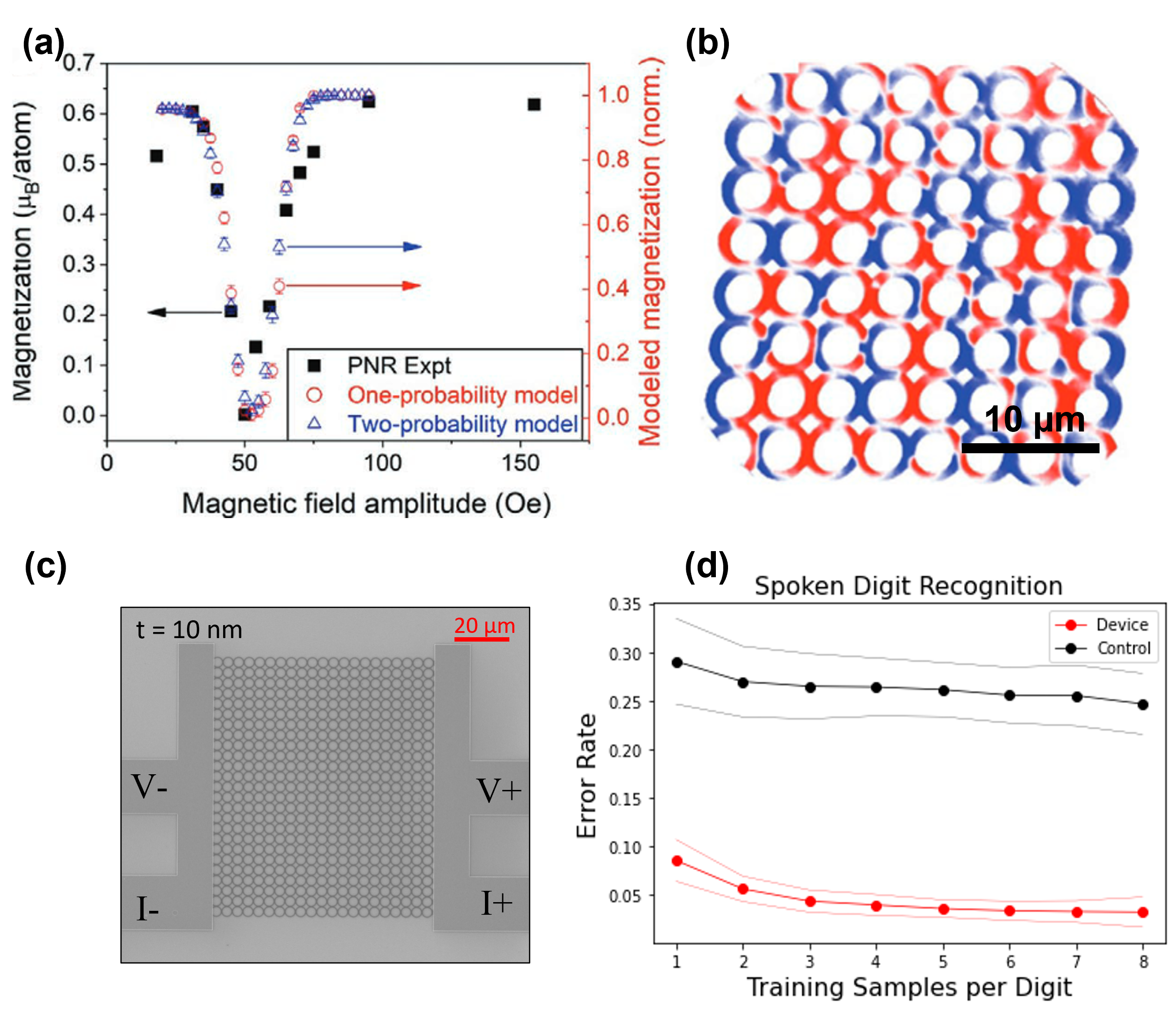}
\caption{(a) The variation of the magnetisation of the ring array with rotating magnetic field amplitude showing a strongly nonlinear response. (b) An X-ray photoemission microscopy image of the ring array showing the different magnetic states existing at an intermediate rotating field amplitude. The arrows indicate the direction of the magnetisation sensitivity. (c) The variation of state populations in the array with number of field rotations for different rotating field amplitudes. The very different evolution for different fields is indicative of the 'fading memory' in the system. Taken from [\onlinecite{dawidek2021dynamically}]}\label{fig23}
\end{figure*}

We shall focus on proposals and demonstrations of RC which use DW dynamics. Dawidek  et al. \cite{dawidek2021dynamically} (Type:Expt) showed that the stochastic nature of DW pinning and depinning in arrays of IM permalloy nanorings are suitable for RC. The arrays were driven by rotating magnetic fields with the DW behaviour in the array changing according to the field strength. Low-strength magnetic fields left DWs pinned at ring junctions, while strong fields caused all DWs to pass through junctions. The stochastic depinning resulting from intermediate fields gave rise to DW annihilation, from a moving DW meeting a DW pinned at a junction, which leads to the formation of magnetic vortex states in some nanorings. However, DWs subsequently moving through a junction within a vortex-state ring repopulated the nanoring with two DWs. A dynamic equilibrium between these DW loss and gain phenomena led to the array magnetisation (via the DW population) becoming a field-dependent emergent property of the ring arrays, as determined by polarised-neutron magnetometry (Figure \ref{fig23} (a)). Imaging array magnetisation using X-ray photoelectron emission microscopy (X-PEEM) revealed a changing and rich magnetic state population in these arrays (Figure \ref{fig23} (b)). The authors also simulated machine learning of spoken digit recognition with these arrays and obtained $99.4\%$ accuracy for classification accuracy for a single speaker and upto $89\%$ accuracy for eight speakers. Subsequently, Vidamour et al. \cite{vidamour2022quantifying} (Type:Sim) used a phenomenological model of the ring array system to optimise the dynamics and performance of these arrays for different classification tasks by tuning the scaling and input rate of data into the reservoir. They used task agnostic metrics for quantifying the determine the capabilities of these arrays for computation and showed the association of these metrics with performance in different tasks.

Vidamour et al. \cite{vidamour2022reservoir} (Type:Expt) then went on to report the fabrication of electrical devices using these arrays (Figure \ref{fig23} (c)) with these arrays and used their anisotropic magnetoresistance (AMR) response for machine learning. They now experimentally evaluated the task agnostic kernel and generalisation rank metrics for assessing the suitability of the array for RC and used these metrics for identifying the operating parameters of the arrays. They went on to demonstrate the performance of various tasks of varying computational requirements of non-linearity and memory (including signal transformation, spoken digit recognition and nonlinear autoregressive moving average series prediction) with the array and achieved state of the art performance. 

\begin{figure*}
   \centering
       \includegraphics[width=2\columnwidth]{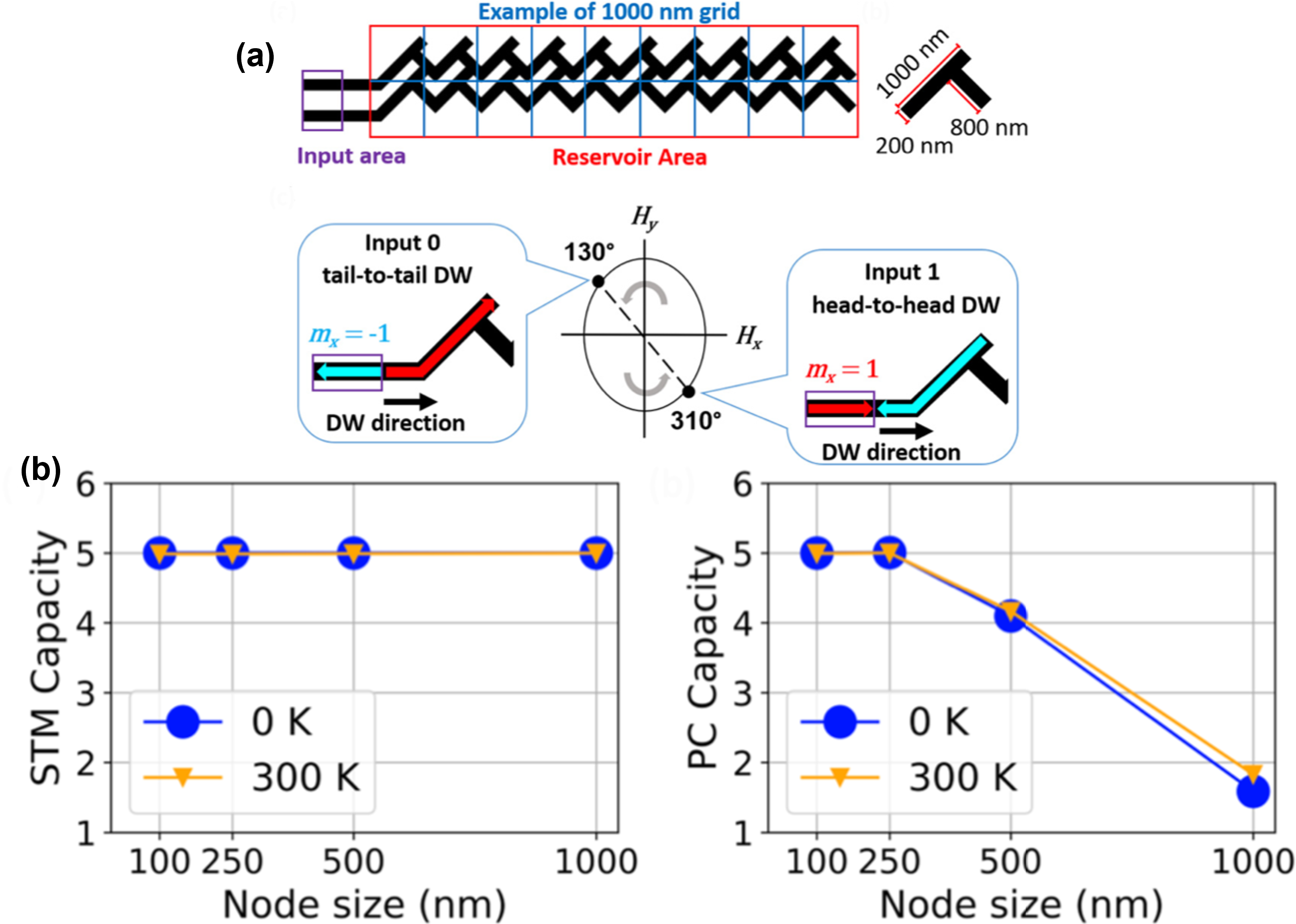}
\caption{(a) A schematic of a grid of $\lambda$-type junctions used for acting as a reservoir. Also shown is the schematic of a single junction with associated dimensions. The method of inputting data into the grid using a clock field (denoted by the ellipse). The states for an input of '0' and of '1' are also shown. (b) The variation of the short term memory (STM) and parity check (PC) of the system at different temperatures. Taken from [\onlinecite{hon2022numerical}]}\label{fig24}
\end{figure*}

Further proposals of RC include that by Ababel et al. \cite{ababei2021neuromorphic} (Type:Sim) who proposed using the oscillations of a single in the potential barrier between two protrusions in an IM Ni nanowires to perform RC. They encoded the data input into the amplitude of a driving magnetic field at $500\,\rm{MHz}$ and used the DW position between two ‘anti-notch’ potential barriers along a nanowire length to represent output. The complex DW propagation dynamics in this simple one-DW system proved capable of performing a number of classification tasks, including sine / square wave differentiation, spoken digits, and handwritten digits. Subsequently, Hon et al. \cite{hon2022numerical} (Type:Sim) performed RC using micromagnetic simulations of an array of nanowires with $\lambda$-shaped junctions (refer to Figure \ref{fig24}). The DW motion, dynamics, and depinning behaviour at junctions under a clocking magnetic field were used to simulate short term memory (STM) and parity check (PC) tasks using RC methodology. The devices would appear to be robust to changes in temperature, with similar performance estimated at 0 and 300 K.

The temporal signal processing utility of RC lends itself to analysing delayed feedback dynamics and a typical example of this are the Mackey-Glass equations \cite{mackey1977oscillation} which are nonlinear, time-delayed differential functions that give rich dynamics, originally envisaged as simulating the dynamics of regulated biological systems. Mackey-Glass oscillators obey these equations and allow oscillatory and chaotic dynamics to be explored in a simple manner. This is particularly exciting for RC as the nonlinear transformations using ‘edge-of-chaos’ dynamics should offer rich and tunable performance. Williame et al. \cite{williame2021magnetic}  (Type:Sim) demonstrated Mackey-Glass oscillator behaviour in simulations of STT-driven DW motion in a nanowire with an elliptical protrusion, although they commented that, in practice, current-driven heating of the device and fabrication uncertainties might affect device performance.

These proposals and demonstrations show that DW dynamics in nanowires devices possess the capabilities required for RC in a variety of tasks. Future work in magnetic implementations of RC will no doubt see a wider exploration of suitable systems, and control and tuning of these systems as well as further demonstrations of machine learning in a variety of tasks and environments.

\subsection{Other applications}\label{S4-4}

The global market for magnetic field sensing was USD $\$2.21$ billion in 2018 and projected to rise to $\$4.22$ billion by 2026 \cite{magneticsensorsmarketresearch2022}. Although this is dominated by Hall sensors, field sensing is a common application of magnetic materials \cite{lenz2006magnetic} and magnetoresistive sensors made up almost 15\% of the global market in 2018. Some proposals of magnetic field sensing have involved DW devices.

Wolfe et al. \cite{wolfe1991fiber} (Type:Expt) proposed a sensor based on the Faraday effect in an IM magnetic garnet thin film ridged waveguide. When a single DW, created using a gradient magnetic field, crosses the path of light incident on the film, the Faraday rotation changes and thus the field causing the magnetisation change in the film may be detected. Diegel et al. \cite{diegel2009new} suggested a four-bit DW-based multi-turn counter, primarily for automotive and industrial applications. The sensor layout consists of four rectangular-shaped spirals, two winding clockwise and two anti-clockwise, of soft ferromagnetic nanowire multilayers that exhibit GMR. Electrical  connections formed a half-bridge configuration that gave stability against temperature variation. The spiral centres had DW injection pads that acted as a source or sink of DWs in the magnetic free layer, depending on the rotation direction of the external magnetic field. This gave opposite sensitivity in the two spiral pairs to the sense of field rotation and caused either an increase or decrease in each spiral’s free-layer DW population. The GMR arrangement meant that these differences could be measured easily to determine the net whole number of turns of magnetic field in either rotation direction.

DWs have also been used for temperature sensing. Klingbeil et al. \cite{klingbeil2021sensing} (Type:Expt) considered meandering DW formation in Bi-substituted rare earth iron garnet films and measured out-of-plane hysteresis loops using a Faraday effect based setup. They monitored the change in the overall magneto-optical response, the DW nucleation fields and the magneto-optical susceptibility (the derivative of the magneto-optical response with field) as temperature was varied. Using appropriate calibration, they could monitor changes in temperature over a $20-140\,^{\circ}\rm{C}$ range with an accuracy of $0.1\,^{\circ}\rm{C}$ and with a temperature drift of better than $0.15\,^{\circ}\rm{C}$.

The stray fields from DWs have also been used to control secondary systems. In particular, there have been several proposals and demonstrations of controlling magnetic nanoparticles (MNPs) for biological applications. Vavassori et al. \cite{vavassori2008domain} (Type:Expt) considered a micron-sized square ring of IP permalloy with a DW positioned at diagonally opposite corners. When a superparamagnetic bead was also positioned at one of the corners and the corresponding DW was displaced using a magnetic field, a dipole moment was induced in the bead and the stray field due to this caused the field required to displace the DW to increase. They simulated this shift in field and also measured a $12\,\rm{Oe}$ shift in the AMR response of the square ring when beads were dispersed on it compared to when they weren't. Furthermore, simulations showed that by reducing the width of the square ring, the shift in field could be increased. Bryan et al. \cite{bryan2010effect} (Type:Sim) used finite element micromagnetic simulations to model the magnetostatic interactions between a superparamagnetic bead and a H2H DW in a nanowire and observed the behaviour of the DW around the stationery bead. They then used analytical formulations to model how the hydrodynamic drag on the bead affects the DW movement. They found that that the DW was pinned around the superparamagnetic bead below a threshold magnetic field which was proportional to the bead diameter squared. Even from small beads, this effect is sufficient to reduce the domain wall mobility by five orders of magnitude. 

\begin{figure*}
   \centering
       \includegraphics[width=2\columnwidth]{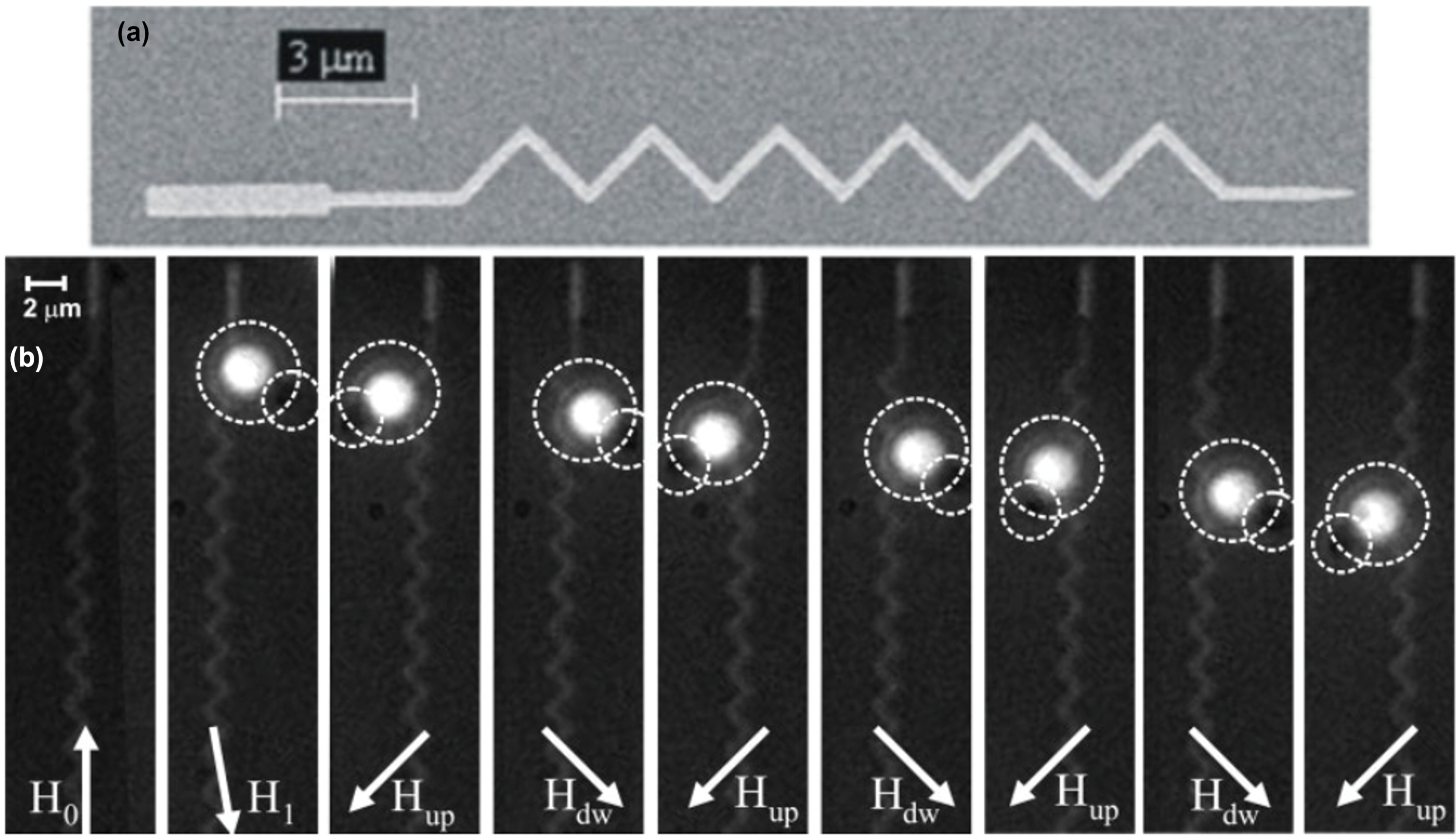}
\caption{(a) A zig-zag wire structure which was used to position DWs and superparamagnetic beads with magnetic fields. (b) Optical images showing the shifting of protein coated magnetic beads by controlling DW movement in the nanowires structure. Taken from [\onlinecite{donolato2010chip}]}.\label{fig25}
\end{figure*}

Shortly afterwards, Donolato et al. \cite{donolato2010chip} (Type:Expt) controlled the movement of protein carrying beads of different sizes exploiting the positioning of DWs in zig-zag wires of permalloy using pulsed magnetic fields (refer to Figure \ref{fig25}). The beads tended to follow the DWs with some drift as the speed of DW motion is considerably faster than the maximum measured bead velocity of $15\,\mu\rm{m/s}$. The modification of the field pulse sequence was used to change the direction of bead movement as well as hold the bead at the DW constriction for analysis (refer to Figure \ref{fig25}). The authors also demonstrated continuous control of bead movement by considering DWs in a ring structure and applyign rotating magnetic fields. Small field steps of $1\,^{\circ}$ were used to obtain fine control of nanoparticle positioning with an accuracy of $100\,\rm{nm}$. Bryan et al. \cite{bryan2010switchable} (Type:Expt) considered different types of nanowires (single planar, two perpendicular and curved) and simulated traps in them to identify the trapping stray field produced. They went on to fabricate these wire and used X-ray imaging to identify that the beads were trapped in the nanowire corners where the DWs were pinned. Subsequently, Rapoport et al. \cite{rapoport2017architecture} considered a closed loop curvilinear permalloy track and simulated the motion of vortex DWs through curvilinear junctions to find that the DWs actually split into a H2H and T2T DW (with opposite stray field polarities) upon exiting the junction. They used these DWs to move a superparamagnetic bead at a junction and to sort beads of different sizes.

Similar concepts have also been extended to controlling the movement of atomic size particles. Allwood et al. \cite{allwood2006mobile} (Type:Sim) used finite element micromagnetic simulations to show that above a DW in a nanowire, a magnetic field trap for a single $^{87}\rm{Rb}$ atom could be formed. Furthermore, they expected that the position of this trap could be modified by applying moving the DW. Subsequently, West et al. \cite{west2012realization} considered an array of undulating IM permalloy nanowires and nucleated DWs in them at remanance after saturating the wires orthogonal to the their length. The resulting large array of tiny magnets worked as a magnetic mirror which could reflect $^{87}\rm{Rb}$ atoms via the fringing field of the magnets. They could switch the magnets on and off reproducibly by switching the magnetisation direction of the array of nanowires.

These studies motivate the use of DW based devices as novel magnetic field, temperature and atomic particle sensors. While they might still be far from commercial application, the versatility of DW dynamics to respond to a variety of stimuli caused by different agents offers new insight into their potential uses.
\section{Future outlooks}\label{S5}

Magnetic DWs have rich and varied behaviour and a wide variety of applications have been proposed utilising them. Here we have reviewed the various types of DWs in magnetic thin films and nanowires with both in-plane and out-of-plane anisotropy. We have reviewed the methods to create and manipulate DW dynamics with a variety of methods and highlighted how rich the behavioural and stimulus space of DWs is in for potential applications. We have also looked at different applications that have been suggested using them and reviewed the state of the art in them.



The state of the art with CMOS based devices for Boolean memories and computation is capable of producing $7\,\rm{nm}$ transistor devices which are facing issues of repeatable device manufacturing and performance \cite{radamson2020state}. In order to be a viable alternative for CMOS devices, DW based devices thus need to be studied and tailored to operate at similar sizes and operate at high speeds with low operational power and latency. Over the last couple of decades, one of the biggest obstacles inhibiting DW proposals for devices from hitting the market in a large scale has been that current densities required for applications involving Oersted field or STT are typically too large for technological feasibility. The current densities required for STT have been declining and with the advent of SOT, there is encouraging signs that we will soon reach reasonable current densities for technological realisations. The material science research required for identifying candidates for high SOT efficiency is key to making more progress in this area. Furthermore, one of the challenges with SOT based devices is that it is still quite sensitive to fabrication conditions and interface qualities. This is a significant step that has to be overcome for reproducible and reliable devices. Another significant problem for Boolean memory and computing applications stochasticity of DW behaviour especially as the magnetic element size is decreased for higher scalability and memory density. Again, material science seems to be important here and the key is to identify material systems that have energy barriers for thermal driven magnetisation changes are high enough that reduction in storage or processing bit size is possible.

However, the stochasticity of DWs lend themselves naturally to non-Boolean applications as has been discussed in Section. \ref{S4-3}. However, before commercial realisation, some important points must be addressed. In the approach where neural network components (like neurons and synapses) are being realised with various DW devices, larger networks and more advanced functionalities need to be demonstrated if these devices are to compete with other existing or proposed technological implementations of NC. Furthermore the issue of reproducability and robustness is something that has not been actively pursued in the community and this is going to be important for technological realisation. In the other forms of unconventional computing that we discussed (like reservoir computing), one of the key pursuits should be to identify applications and tasks which can't be easily (or economically) performed with existing CMOS based systems. This will require demonstrating advantages of the DW-based implementations in terms of niche applications, energy throughput or other relevant metric. We are quite excited to see how this research field progresses and where it take us.

\begin{acknowledgments}
GV, DAA and TJH acknowledge funding from the Horizon 2020 FET-Open SpinEngine (Agreement no 861618), the EPSRC MARCH project EP/V006339/1, the Leverhulme grant RPG-2019-97 and the EPSRC project EP/S009647/1. We would also like to acknowledge C. Swindells, I. Vidamour and A. Welbourne for discussions.
\end{acknowledgments}

\section*{References}
\bibliography{GV-mega-bibliography}

\end{document}